\documentclass[11pt,onecolumn,english,preprintnumbers,amsmath,amssymb,floatfix,nofootinbib]{revtex4}
\usepackage{tikz,xcolor}
\usepackage[colorlinks = true,  
linkcolor = blue,
urlcolor  = blue,
citecolor = blue,
anchorcolor = blue]{hyperref}

\definecolor{lime}{HTML}{A6CE39}
\DeclareRobustCommand{\orcidicon}{%
	\begin{tikzpicture}
		\draw[lime, fill=lime] (0,0) 
		circle [radius=0.16] 
		node[white] {{\fontfamily{qag}\selectfont \tiny ID}};
		\draw[white, fill=white] (-0.0625,0.095) 
		circle [radius=0.007];
	\end{tikzpicture}
	\hspace{-2mm}
}

\foreach \x in {A, ..., Z}{%
	\expandafter\xdef\csname orcid\x\endcsname{\noexpand\href{https://orcid.org/\csname orcidauthor\x\endcsname}{\noexpand\orcidicon}}
}

\usepackage[T1]{fontenc}
\usepackage[latin9]{inputenc}
\usepackage{color}
\usepackage{array}
\usepackage{amstext}
\usepackage{graphicx}
\usepackage{esint}
\usepackage{rotating}
\usepackage{appendix}
\usepackage{float}
\usepackage{subcaption}
\usepackage{amsmath}
\usepackage{hyperref}
\usepackage{booktabs}   
\usepackage{graphicx}   
\usepackage{array} 
\usepackage{adjustbox}

\usepackage[font=small,labelfont=bf]{caption}
\usepackage{xcolor}
\usepackage{ulem}

\makeatletter

\@ifundefined{textcolor}{}
{%
	\definecolor{BLACK}{gray}{0}
	\definecolor{WHITE}{gray}{1}
	\definecolor{RED}{rgb}{1,0,0}
	\definecolor{GREEN}{rgb}{0,1,0}
	\definecolor{BLUE}{rgb}{0,0,1}
	\definecolor{CYAN}{cmyk}{1,0,0,0}
	\definecolor{MAGENTA}{cmyk}{0,1,0,0}
	\definecolor{YELLOW}{cmyk}{0,0,1,0}
}

\@ifundefined{definecolor}
{\usepackage{color}}{}
\@ifundefined{definecolor}
{\usepackage{color}}{}
\makeatother
\usepackage{babel}

\newcommand{\beq}{\begin{equation}}
	\newcommand{\eeq}{\end{equation}}

\def\Re{{\cal R \mskip-4mu \lower.1ex \hbox{\it e}\,}}
\def\Im{{\cal I \mskip-5mu \lower.1ex \hbox{\it m}\,}}

\def\tev{\,{\ifmmode\mathrm {TeV}\else TeV\fi}}
\def\gev{\,{\ifmmode\mathrm {GeV}\else GeV\fi}}
\def\mev{\,{\ifmmode\mathrm {MeV}\else MeV\fi}}
\def\to{\rightarrow}

\begin{document}

\title{Probing aQGCs from dimension-eight operators via neutral triboson production at muon colliders}

\author{Parimah Hosseini$^{1}$\orcidA}
\email{parimah.hosseini@ut.ac.ir}

\author{Sara Khatibi$^{1}$ \orcidC}
\email{sara.khatibi@ut.ac.ir}

\author{Hananeh Sadat Modirzadeh$^{2,3}$\orcidB}
\email{hsmodir@ipm.ir}

\affiliation{
$^{1}$ Department of Physics, University of Tehran, North Karegar Ave., Tehran 14395-547, Iran \\
$^{2}$ Department of Physics, Sharif University of Technology, Tehran, Iran \\
$^{3}$ School of Astronomy, Institute for Research in Fundamental Sciences (IPM), Tehran 19395-5531, Iran
}

\date{\today}

\begin{abstract}
We investigate the sensitivity of a future high-energy muon collider to neutral anomalous quartic gauge couplings (aQGCs) 
through neutral triboson production processes, specifically $\mu^+\mu^- \to \gamma\gamma\gamma$, $\gamma\gamma Z$, 
$\gamma ZZ$, and $ZZZ$. Within the Standard Model Effective Field Theory (SMEFT) framework, we focus on genuine dimension-eight 
transverse operators ($\mathcal{O}_{T_0}$--$\mathcal{O}_{T_9}$). The analysis is performed at prospective 
center-of-mass energies of $\sqrt{s} = 3\text{ TeV}$ and $10\text{ TeV}$ with integrated luminosities of 
$1\text{ ab}^{-1}$ and $10\text{ ab}^{-1}$, respectively. Employing an optimized cut-and-count analysis 
and a $\chi^2$ statistical framework incorporating a $5\%$ systematic uncertainty, we derive projected 
95\% confidence level bounds on the corresponding Wilson coefficients. Our results demonstrate that the 
triphoton channel yields the most stringent constraints, particularly for the $f_{T_8}$ and $f_{T_9}$ 
coefficients, while channels involving $Z$ bosons exhibit reduced sensitivity owing to leptonic branching 
ratio suppressions. Furthermore, we report constraints on the $f_{T_3}$ and $f_{T_4}$ coefficients, 
established here for the first time in this neutral triboson context, highlighting the powerful 
physics reach of multi-TeV muon colliders in probing electroweak gauge sector interactions.
\end{abstract}

\maketitle	
	
	\newpage
	\section{Introduction} \label{intro}

The Standard Model (SM) is currently the most comprehensive theoretical framework describing elementary particles and their interactions. Despite its remarkable experimental success and precise predictions, it fails to explain several phenomena. These limitations indicate that the SM alone cannot be the complete picture and motivate the search for physics beyond the Standard Model (BSM). One promising probe of BSM physics is the study of anomalous gauge boson couplings. Anomalous triple gauge couplings (aTGCs) \cite{Ellis:2014dva, Ellis:2020ljj} and anomalous quartic gauge couplings (aQGCs) \cite{ATLAS:2026wew, CMS:2026job} provide a means of probing BSM physics through deviations from the SM predictions for gauge boson interactions by either modifying interaction vertices currently present in the SM or introducing entirely new vertices that are absent in the SM \cite{Degrande:2018fog}. For instance, interaction vertices involving four neutral gauge bosons---i.e., $\gamma\gamma\gamma\gamma$, $Z\gamma\gamma\gamma$, $ZZ\gamma\gamma$, $ZZZ\gamma$, and $ZZZZ$---are forbidden at tree level in the SM, making neutral aQGCs an interesting probe of new physics. 

Anomalous gauge couplings can be studied using anomalous-coupling parametrizations \cite{Belanger:1992qh, Gupta:2011be} or within a gauge-invariant effective field theory (EFT) framework, such as the Standard Model Effective Field Theory (SMEFT) \cite{Degrande:2012wf, Brivio:2017vri, Celada:2024cxw}. In the former approach, which is now largely superseded and does not preserve the SM gauge symmetry, anomalous interactions are introduced directly through effective interaction vertices. In the latter approach, deviations from the SM are described by introducing operators with higher mass dimension, which are constructed from the SM fields and preserve the $SU(3)_C \otimes SU(2)_L \otimes U(1)_Y$ symmetry. The coupling strengths in an EFT Lagrangian are commonly referred to as Wilson coefficients. Some dimension-six and dimension-eight operators contribute to aTGCs and aQGCs, so the SMEFT framework can be used to derive constraints on the Wilson coefficients of these operators \cite{Falkowski:2016cxu, Remmen:2019cyz}. Dimension-six operators can generate correlated aTGCs and aQGCs, whose Wilson coefficients are more commonly constrained via the aTGC part. Independent aQGCs first appear at dimension eight and are generated by \textit{genuine} quartic gauge coupling (QGC) operators \cite{Eboli:2006wa}. The transverse operators $\mathcal{O}_{T_{i}}$, a subclass of genuine dimension-eight operators, contribute to all aforementioned neutral quartic vertices \cite{Almeida:2020ylr}.

In practice, aTGCs and aQGCs are commonly studied through the production of two electroweak gauge bosons (diboson production) \cite{CMS:2020gtj, Degrande:2023iob, ElFaham:2024uop, Gillies:2024mqp}, vector boson scattering (VBS) \cite{CMS:2020ioi, Yang:2022fhw} and processes involving the production of three electroweak gauge bosons (triboson production). Triboson production is a rare process that provides a means of probing quartic gauge interaction vertices directly. Various triboson production channels have been observed and measured at the Large Hadron Collider (LHC) by the ATLAS and CMS collaborations \cite{ATLAS:2017lpx, CMS:2020hjs, CMS:2021jji, ATLAS:2022wmu, ATLAS:2024nab, ATLAS:2026zgd, CMS:2026vhr}. The sensitivity of triboson production to aQGCs makes it a valuable tool for probing new physics. Probing aQGCs via triboson production channels within the SMEFT framework has been explored at various proposed future hadron and lepton colliders, including FCC-hh \cite{Senol:2021wza, Senol:2023fpa, Yilmaz:2026afl}, CLIC \cite{Ari:2021rmx}, and muon colliders \cite{Yang:2020rjt, Zhang:2023yfg, Dong:2023nir}. 

Among proposed future colliders, a high-energy muon collider offers a well-suited environment to explore electroweak interactions with high precision. Since muons are elementary particles, the entire beam energy is accessible in the hard collision, and since they are much heavier than electrons, they lose far less energy to synchrotron radiation, making a multi-TeV muon collider an effective machine for probing high-energy interactions and BSM searches. Moreover, QCD background is significantly suppressed at muon colliders compared to hadron colliders, as the purely leptonic initial state includes no incoming quarks or gluons at leading order (LO) \cite{Shiltsev:2019rfl, AlAli:2021let, Aime:2022flm}.

In this paper, we investigate the contributions of genuine dimension-eight operators of the $\mathcal{O}_{T_{i}}$ class to the fully neutral aQGCs $\gamma\gamma\gamma\gamma$, $Z\gamma\gamma\gamma$, $ZZ\gamma\gamma$, $ZZZ\gamma$, and $ZZZZ$. We study $\gamma\gamma\gamma$, $\gamma\gamma Z$, $\gamma ZZ$, and $ZZZ$ production at muon colliders with $\sqrt{s} = 3~\text{TeV}$ and $10~\text{TeV}$, with the $Z$ bosons decaying to charged leptons ($e$ and $\mu$). We perform a cut-based analysis and use a standard $\chi^{2}$ analysis to extract 95\% confidence level (CL) bounds on the Wilson coefficients, assuming a 5\% systematic uncertainty.

This paper is organized as follows. In Section~\ref{sec:model} we provide a detailed description of the theoretical framework underlying our analysis. Section~\ref{sec:analysis} contains the details of our simulation, including Monte Carlo event generation, detector simulation, and signal and background separation techniques. We present our results and the constraints derived on the considered Wilson coefficients in Section~\ref{sec:result}. Finally, we summarize our findings in Section~\ref{sec:summary}, including a comparison with existing constraints in the literature.

	\section{Theoretical Framework}\label{sec:model}
	
To date, no extension of the SM has been experimentally confirmed, which motivates exploring potential new physics signatures through model-independent approaches. EFTs provide a framework in which low-energy effects of BSM physics can be parametrized. In particular, SMEFT is an EFT formulated using the same field content as the SM, preserving its symmetries. Within this framework, BSM effects are parametrized by higher-dimensional operators added to the SM Lagrangian. The full SMEFT Lagrangian can be written as \cite{Falkowski:2023hsg, Degrande:2012wf}
\begin{equation}
\mathcal{L}_{\text{SMEFT}}
=
\mathcal{L}_{\text{SM}}
+\sum_{d>4}\sum_i
\frac{C_i^{(d)}}{\Lambda^{d-4}}
\mathcal{O}_i^{(d)},
\label{eq:SMEFT_Lagrangian}
\end{equation}
where $\mathcal{L}_{\text{SM}}$ is the SM Lagrangian, $\mathcal{O}_i^{(d)}$ denotes a gauge-invariant operator of mass dimension $d$ constructed from the SM fields, ${C_i^{(d)}}$ is the corresponding dimensionless Wilson coefficient, and $\Lambda$ is the new-physics scale. The leading higher-dimensional contribution occurs at dimension five and is represented by the Weinberg operator, which can account for Majorana neutrino masses. Dimension-six and dimension-eight operators introduce additional contributions to the SM processes and have been extensively studied in the literature. One of their many applications is the study of anomalous gauge boson couplings \cite{Etesami:2016rwu, Tizchang:2020tqs}. In the SM, the self-interactions of gauge bosons are determined by the non-Abelian structure of the electroweak gauge group: the triple and quartic vertices involving charged gauge bosons are present at tree level and have a completely specified strength and Lorentz structure. However, vertices involving only neutral gauge bosons are absent at tree level. In this context, aTGCs and aQGCs refer to deviations from the SM predictions, including modified charged-boson vertices or new vertices forbidden in the SM at tree level. Such deviations can be induced by new physics at a scale $\Lambda$ beyond the reach of current experiments. 

In the SMEFT approach, dimension-six operators induce aTGCs together with correlated aQGCs, whereas aQGCs without any associated aTGCs first arise at dimension eight through genuine QGC operators \cite{Eboli:2006wa}. In this work, we focus exclusively on the contributions of genuine dimension-eight operators to the aQGCs, and assume that all dimension-six operator coefficients vanish. This truncation assumes that the EFT expansion remains valid at the energies probed in our analysis, with contributions from higher-dimensional operators sufficiently suppressed. The resulting constraints should therefore be interpreted within the assumed dimension-eight framework.

A basis of genuine QGC operators was first introduced in Ref.~\cite{Eboli:2006wa} and was later updated in Refs.~\cite{Eboli:2016kko, Almeida:2020ylr}. The operators in this basis generate various quartic gauge interactions and, depending on their field content, associated interactions involving the Higgs boson. The operators under consideration are even under both charge conjugation ($C$) and parity ($P$). A complete list of independent genuine QGC operators that are both $C$-even and $P$-even is presented in Ref.~\cite{Almeida:2020ylr}. (Ref.~\cite{Almeida:2020ylr} does not include two independent operators that are $C$-odd and $P$-odd but $CP$-even, which we do not consider in this work either. A complete list of independent QGC operators, specifying their CP properties, is presented in Ref.~\cite{Durieux:2024zrg}. See also Refs.~\cite{Murphy:2020rsh, Li:2020gnx}.) In this paper, we adopt the notation presented in Ref.~\cite{Almeida:2020ylr}. 

The genuine dimension-eight operators can be classified into three categories: scalar (or longitudinal) operators $\mathcal{O}_{S_{i}}$ that contain only the covariant derivatives of the scalar field, mixed operators $\mathcal{O}_{M_{i}}$ that contain both covariant derivatives of the scalar field and the gauge field strength tensors, and transverse operators $\mathcal{O}_{T_{i}}$ that contain only the field strength tensors. The effective Lagrangian is then \cite{Almeida:2020ylr}
\begin{equation}
    \mathcal{L} = \mathcal{L}_{\mathrm{SM}} + \mathcal{L}_{8},
    \label{eq:total_Lagrangian}
\end{equation}
in which the dimension-eight contribution is
\begin{equation}
\mathcal{L}_{8} =
\sum_{i=0}^{2} \frac{f_{S_{i}}}{\Lambda^4}\mathcal{O}_{S_{i}}
+
\sum_{i=0,1,2,3,4,5,7} \frac{f_{M_{i}}}{\Lambda^4}\mathcal{O}_{M_{i}}
+
\sum_{i=0}^{9} \frac{f_{T_{i}}}{\Lambda^4}\mathcal{O}_{T_{i}},
\label{eq:dimEight_Lagrangian}
\end{equation}
where $f_{S_{i}}$, $f_{M_{i}}$, and $f_{T_{i}}$ denote the dimensionless Wilson coefficients associated with their corresponding operators, and $\Lambda$ represents the energy scale of new physics. Note that $f_{i}$ in Eq.~\eqref{eq:dimEight_Lagrangian} corresponds to $C_{i}^{(8)}$ in Eq.~\eqref{eq:SMEFT_Lagrangian}. It should be noted that $\mathcal{O}_{M_{6}}$ is excluded from Eq.~\eqref{eq:dimEight_Lagrangian}, since it was found to be redundant \cite{Eboli:2016kko}. Note also that $\mathcal{O}_{S_{2}}$, which was initially absent in Ref.~\cite{Eboli:2006wa} and in several later works but was included in Ref.~\cite{Eboli:2016kko}, is also present in Eq.~\eqref{eq:dimEight_Lagrangian}. 

In the early studies of aQGCs, the $\mathcal{O}_{T_{3}}$ and $\mathcal{O}_{T_{4}}$ operators were absent in the dimension-eight operator basis \cite{Eboli:2006wa, Eboli:2016kko}, whereas later studies adopted the complete basis comprising all ten $\mathcal{O}_{T_{i}}$ operators \cite{Almeida:2020ylr, Durieux:2024zrg}. In this work, we adopt the complete operator basis consistent with Ref.~\cite{Almeida:2020ylr}, including the $\mathcal{O}_{T_{3}}$ and $\mathcal{O}_{T_{4}}$ operators, which are also considered in more recent analyses \cite{CMS:2025dbm, CMS:2026job, ATLAS:2026hrg}. One can verify that $\mathcal{O}_{T_{3}}$ modifies $WWWW, WWZZ, WWZ\gamma, WW\gamma\gamma, ZZZZ, ZZZ\gamma, ZZ\gamma\gamma, Z\gamma\gamma\gamma, \gamma\gamma\gamma\gamma$ vertices. Similarly, the operator $\mathcal{O}_{T_{4}}$ also contributes to the same vertices excluding $WWWW$. The contributions of the genuine dimension-eight operators to aQGCs are summarized in Table~\ref{tab:QGC_operators}.
\begin{table}[htbp]
    \centering
    \caption{Contributions of the genuine dimension-eight operators to aQGCs are marked with $\checkmark$.}
    \label{tab:QGC_operators}
    \begin{tabular}{|c|c|c|c|c|c|c|c|c|c|}
        \hline
           & $WWWW$ & $WWZZ$ & $ZZZZ$ & $WWZ\gamma$ & $WW\gamma\gamma$ & $ZZZ\gamma$ & $ZZ\gamma\gamma$ & $Z\gamma\gamma\gamma$ & $\gamma\gamma\gamma\gamma$ \\
        \hline
        $\mathcal{O}_{S_{0}}$, $\mathcal{O}_{S_{1}}$, $\mathcal{O}_{S_{2}}$ & $\checkmark$ & $\checkmark$ & $\checkmark$ &  &  &  &  &  &  \\
        \hline
        $\mathcal{O}_{M_{0}}$, $\mathcal{O}_{M_{1}}$, $\mathcal{O}_{M_{7}}$ & $\checkmark$ & $\checkmark$ & $\checkmark$ & $\checkmark$ & $\checkmark$ & $\checkmark$ & $\checkmark$ &  &  \\
        \hline
        $\mathcal{O}_{M_{2}}$, $\mathcal{O}_{M_{3}}$, $\mathcal{O}_{M_{4}}$, $\mathcal{O}_{M_{5}}$ &  & $\checkmark$ & $\checkmark$ & $\checkmark$ & $\checkmark$ & $\checkmark$ & $\checkmark$ &  &  \\
        \hline
        $\mathcal{O}_{T_{0}}$, $\mathcal{O}_{T_{1}}$, $\mathcal{O}_{T_{2}}$, $\mathcal{O}_{T_{3}}$ & $\checkmark$ & $\checkmark$ & $\checkmark$ & $\checkmark$ & $\checkmark$ & $\checkmark$ & $\checkmark$ & $\checkmark$ & $\checkmark$ \\
        \hline
        $\mathcal{O}_{T_{4}}$, $\mathcal{O}_{T_{5}}$, $\mathcal{O}_{T_{6}}$, $\mathcal{O}_{T_{7}}$ &  & $\checkmark$ & $\checkmark$ & $\checkmark$ & $\checkmark$ & $\checkmark$ & $\checkmark$ & $\checkmark$ & $\checkmark$ \\
        \hline
        $\mathcal{O}_{T_{8}}$, $\mathcal{O}_{T_{9}}$ &  &  & $\checkmark$ &  &  & $\checkmark$ & $\checkmark$ & $\checkmark$ & $\checkmark$ \\
        \hline
    \end{tabular}
\end{table}

\begin{figure}[htbp]
    \centering

    \begin{subfigure}[c]{0.34\textwidth}
        \centering
        \includegraphics[width=\linewidth]{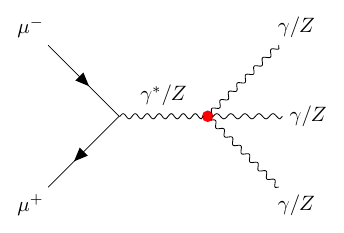}
        \caption{}
        \label{fig:AnomalousCoupling_diagram}
    \end{subfigure}
    \hspace{0.02\textwidth}
    \begin{subfigure}[c]{0.24\textwidth}
        \centering
        \includegraphics[width=\linewidth]{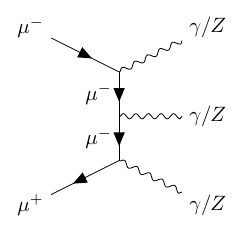}
        \caption{}
        \label{fig:SM_diagram}
    \end{subfigure}
    \begin{subfigure}[c]{0.34\textwidth}
        \centering
        \includegraphics[width=\linewidth]{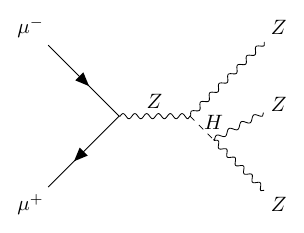}
        \caption{}
        \label{fig:SM_ZZZ_diagram}
    \end{subfigure}

    \caption{Feynman diagrams for neutral triboson production processes at a muon collider: (a) aQGC contributions. The red vertex is induced by $\mathcal{O}_{T_{i}}$ operators; (b) SM contribution; (c) additional SM contribution in the $\mu^{+}\mu^{-}\to ZZZ$ process.}
    \label{fig:process_diagrams}
\end{figure}

Typical Feynman diagrams for neutral triboson production processes at tree level are shown in Figure~\ref{fig:process_diagrams}, 
illustrating both aQGC and SM contributions. At tree level, triboson production with only neutral gauge bosons in the final state 
can contribute only to neutral aQGCs. Since the $\mathcal{O}_{T_{i}}$ operators contribute to all neutral aQGCs, 
we focus exclusively on the transverse operators and derive constraints on the Wilson coefficients of all 
ten $\mathcal{O}_{T_{i}}$ operators. The explicit forms of these operators are \cite{Almeida:2020ylr}
\begin{equation}
\begin{aligned}
\mathcal{O}_{T_{0}} &=
\mathrm{Tr}\left[\hat{W}_{\mu\nu}\hat{W}^{\mu\nu}\right]
\times
\mathrm{Tr}\left[\hat{W}_{\alpha\beta}\hat{W}^{\alpha\beta}\right],
\\
\mathcal{O}_{T_{1}} &=
\mathrm{Tr}\left[\hat{W}_{\alpha\nu}\hat{W}^{\mu\beta}\right]
\times
\mathrm{Tr}\left[\hat{W}_{\mu\beta}\hat{W}^{\alpha\nu}\right],
\\
\mathcal{O}_{T_{2}} &=
\mathrm{Tr}\left[\hat{W}_{\alpha\mu}\hat{W}^{\mu\beta}\right]
\times
\mathrm{Tr}\left[\hat{W}_{\beta\nu}\hat{W}^{\nu\alpha}\right],
\\
\mathcal{O}_{T_{3}} &=
\mathrm{Tr}\left[\hat{W}_{\mu\nu}\hat{W}_{\alpha\beta}\right]
\times
\mathrm{Tr}\left[\hat{W}^{\alpha\nu}\hat{W}^{\mu\beta}\right],
\\
\mathcal{O}_{T_{4}} &=
\mathrm{Tr}\left[\hat{W}_{\mu\nu}\hat{W}_{\alpha\beta}\right]
\times B^{\alpha\nu}B^{\mu\beta},
\\
\mathcal{O}_{T_{5}} &=
\mathrm{Tr}\left[\hat{W}_{\mu\nu}\hat{W}^{\mu\nu}\right]
\times B_{\alpha\beta}B^{\alpha\beta},
\\
\mathcal{O}_{T_{6}} &=
\mathrm{Tr}\left[\hat{W}_{\alpha\nu}\hat{W}^{\mu\beta}\right]
\times B_{\mu\beta}B^{\alpha\nu},
\\
\mathcal{O}_{T_{7}} &=
\mathrm{Tr}\left[\hat{W}_{\alpha\mu}\hat{W}^{\mu\beta}\right]
\times B_{\beta\nu}B^{\nu\alpha},
\\
\mathcal{O}_{T_{8}} &=
B_{\mu\nu}B^{\mu\nu}B_{\alpha\beta}B^{\alpha\beta},
\\
\mathcal{O}_{T_{9}} &=
B_{\alpha\mu}B^{\mu\beta}B_{\beta\nu}B^{\nu\alpha},
\end{aligned}
\label{eq:dimEight_operators}
\end{equation}
where $\hat{W}_{\mu\nu} = \sum_{i=1}^{3}W^i_{\mu\nu}\frac{\sigma^i}{2}$ with $\sigma^i$ being the Pauli matrices, is the $SU(2)_L$ field strength tensor and $B_{\mu \nu}$ is the hypercharge field strength tensor of $U(1)_Y$. 

The total cross~section for each process can be written as 
\begin{equation}\label{eq:total_xsection}
  \sigma_{\mathrm{tot}} = \sigma_{\mathrm{SM}} + (\frac{f}{\Lambda^{4}})\sigma_{\mathrm{int}} + (\frac{f}{\Lambda^{4}})^{2}\sigma_{\mathrm{aQGC}}, 
\end{equation}
where $\sigma_{\mathrm{SM}}$ denotes the SM cross~section, $\sigma_{\mathrm{int}}$ corresponds to the interference between the SM and dimension-eight operators, and $\sigma_{\mathrm{aQGC}}$ represents the pure dimension-eight contribution. In this work, one Wilson coefficient is switched on at a time, while all others are set to zero. The resulting limits therefore ignore cross terms between EFT coefficients and are not the result of a global fit. 

\section{Analysis Strategy}\label{sec:analysis}

In this section, we investigate neutral triboson production processes at a future high-energy muon collider
to probe aQGCs and constrain the genuine dimension-eight 
transverse operators ($\mathcal{O}_{T_0}$--$\mathcal{O}_{T_9}$).
The analysis is performed at two center-of-mass energies, $\sqrt{s} = 3$ and $10~\mathrm{TeV}$, 
assuming integrated luminosities of $1$ and $10~\mathrm{ab}^{-1}$, respectively. Throughout this analysis, 
both the initial-state muon and antimuon beams are taken to be unpolarized. The specific processes considered in this work are listed below:
\begin{align}
\mu^{+} \mu^{-} &\rightarrow \gamma \gamma \gamma, \\
\mu^{+} \mu^{-} &\rightarrow \gamma \gamma Z \rightarrow \gamma \gamma l^{+} l^{-}, \\
\mu^{+} \mu^{-} &\rightarrow \gamma Z Z \rightarrow \gamma l^{+} l^{-} l^{+} l^{-}, \\
\mu^{+} \mu^{-} &\rightarrow Z Z Z \rightarrow e^{+} e^{-} e^{+} e^{-} e^{+} e^{-},
\end{align}
where $l = e, \mu$. For the $ZZZ$ channel, we consider only the electron final state 
and only $\sqrt{s} = 3~\mathrm{TeV}$, since event generation at $10~\mathrm{TeV}$ proved challenging. 
Hence, a more dedicated study of this channel is left to future work.

All relevant contributing processes are taken into account and examined in detail in the subsequent subsections. First, we describe the event-generation framework and evaluate the signal and background cross~sections. Next, we present the event-selection strategies employed to optimize the separation between signal and SM background in each channel separately.
	
\subsection{Event Generation}\label{sec:event}

To model the contributions of the genuine dimension-eight QGC operators discussed in Section~\ref{sec:model}, we utilize the \texttt{QCKM-5-Aug21v2} Universal FeynRules Output (UFO) model~\cite{Eboli:2006wa,Degrande:2011ua}. This UFO model is imported into \textsc{MadGraph5\_aMC@NLO}~\cite{Alwall:2011uj} to generate signal events, where one Wilson coefficient is switched on at a time while all others are set to zero. Relevant SM background samples for each signal process are likewise generated at LO in \textsc{MadGraph5\_aMC@NLO}.

The parton-level events are subsequently passed to \texttt{Pythia 8.3}~\cite{Bierlich:2022pfr} for parton showering and hadronization. Detector effects are then incorporated using \texttt{Delphes 3}~\cite{deFavereau:2013fsa}, with the detector response and object reconstruction modeled utilizing the muon collider detector performance card~\cite{Accettura:2023,MuonColliderDelphesCard}. 

For the event generation, we impose standard default cuts alongside baseline requirements to regulate infrared divergences, demanding a minimum transverse momentum of $p_{T}^{\gamma,l} > 10~\mathrm{GeV}$ for both photons and charged leptons, a pseudorapidity bound of $|\eta_{\gamma,l}| < 2.5$ also for both photons and charged leptons, and a minimum angular separation of $\Delta R > 0.4$ for all photon-photon, photon-lepton, and lepton-lepton pairs, in accordance with the default generation-level cuts of \textsc{MadGraph5\_aMC@NLO}. These selection cuts are consistently applied to all final-state decay products. Hereafter, these requirements are referred to as generation-level cuts.

\subsection{Signal and Background Separation}\label{sec:separation}

To suppress SM background processes, we analyze kinematic observables that distinguish potential signal from background events. Based on these features, we apply a cut-and-count analysis strategy centered on the most sensitive kinematic observable to extract bounds on the Wilson coefficients. In the following subsections, we describe the implementation of this method for each triboson production channel separately.

\subsubsection{The $\gamma\gamma\gamma$ Final State}

The first process under consideration features a pure triphoton final state ($\gamma\gamma\gamma$). Due to the presence of three massless final-state photons and the absence of massive vector boson decay branchings, this channel offers high signal yields and enhanced sensitivity to neutral aQGCs, thereby allowing us to derive particularly stringent constraints on the corresponding Wilson coefficients. Although the triphoton final state at a muon collider has been previously investigated in Refs.~\cite{Yang:2020rjt, Zhang:2023yfg, Dong:2023nir}, we include a re-evaluation of this process in the present study to provide a complete and unified treatment of all neutral triboson final states within the same framework. Since all ten transverse operators $\mathcal{O}_{T_i}$ generate neutral aQGCs, each operator contributes to this final state. The cross~sections obtained from \textsc{MadGraph5\_aMC@NLO} simulations are parametrized as quadratic fit functions of the Wilson coefficients at center-of-mass energies of $\sqrt{s} = 3$ and $10~\mathrm{TeV}$, as summarized in Table~\ref{tab:triphtonfit}. The primary SM background contributions to the triphoton final state originate from irreducible $\gamma\gamma\gamma$ production as well as the $\gamma\gamma\gamma\gamma$ process, both of which are explicitly accounted for in this analysis.
	
\begin{table}[ht!]
\centering
\caption{Quadratic fits to the cross~sections as functions of the effective coupling parameter $\lambda = f_{T_i}/\Lambda^4$ for the $\gamma\gamma\gamma$ final state at center-of-mass energies of $\sqrt{s} = 3$ and $10~\mathrm{TeV}$.}
\scriptsize
\renewcommand{\arraystretch}{1.1}
\setlength{\tabcolsep}{3pt}

\begin{adjustbox}{max width=\textwidth}
\begin{tabular}{|c|c|c|}
\hline
Parameter $\lambda$ ($\mathrm{GeV}^{-4}$) &
Quadratic functions for $\gamma\gamma\gamma$ final state at $3~\mathrm{TeV}$ (pb) &
Quadratic functions for $\gamma\gamma\gamma$ final state at $10~\mathrm{TeV}$ (pb) \\
\hline
$f_{T_0}/\Lambda^4(f_{T_1}/\Lambda^4$) &
$\sigma(\lambda)= 5.95 \times10^{-3} + 3.50 \times10^{8}\lambda + 1.37 \times10^{21}\lambda^{2}$ &
$\sigma(\lambda)= 7.09  \times10^{-4} - 9.25 \times10^{8}\lambda+ 1.81 \times10^{24}\lambda^{2}$ \\
\hline
$f_{T_2}/\Lambda^4(f_{T_3}/\Lambda^4$) &
$\sigma(\lambda)= 5.95 \times10^{-3} + 3.03 \times10^{8}\lambda + 3.40 \times10^{20}\lambda^{2}$ &
$\sigma(\lambda)= 7.09  \times10^{-4} + 1.95 \times10^{9}\lambda + 4.55 \times10^{23} \lambda^{2}$ \\
\hline
$f_{T_4}/\Lambda^4(f_{T_7}/\Lambda^4$) &
$\sigma(\lambda)= 5.95 \times10^{-3} + 1.84 \times10^{9}\lambda+ 7.37 \times10^{21}\lambda^{2}$ &
$\sigma(\lambda)= 7.09  \times10^{-4} - 2.38 \times10^{10}\lambda + 1.02 \times10^{25} \lambda^{2}$ \\
\hline
$f_{T_5}/\Lambda^4(f_{T_6}/\Lambda^4$) &
$\sigma(\lambda)= 5.95 \times10^{-3} + 2.45 \times10^{9}\lambda+ 2.92 \times10^{22}\lambda^{2}$ &
$\sigma(\lambda)= 7.09  \times10^{-4} - 1.48 \times10^{11}\lambda + 4.04 \times10^{25} \lambda^{2}$ \\
\hline
$f_{T_8}/\Lambda^4$ &
$\sigma(\lambda)= 5.95 \times10^{-3} + 2.85 \times10^{10}\lambda + 1.11 \times10^{24}\lambda^{2}$ &
$\sigma(\lambda)= 7.09  \times10^{-4} - 4.27 \times10^{12}\lambda + 1.56 \times10^{27} \lambda^{2}$ \\
\hline
$f_{T_9}/\Lambda^4$ &
$\sigma(\lambda)= 5.95 \times10^{-3} + 1.13 \times10^{10}\lambda + 2.85  \times10^{23}\lambda^{2}$ &
$\sigma(\lambda)= 7.09  \times10^{-4} - 2.12 \times10^{12}\lambda + 3.94 \times10^{26} \lambda^{2}$ \\
\hline
\end{tabular}
\end{adjustbox}
\label{tab:triphtonfit}
\end{table}

\begin{figure*}[htbp]
\centering
\includegraphics[width=0.45\textwidth]{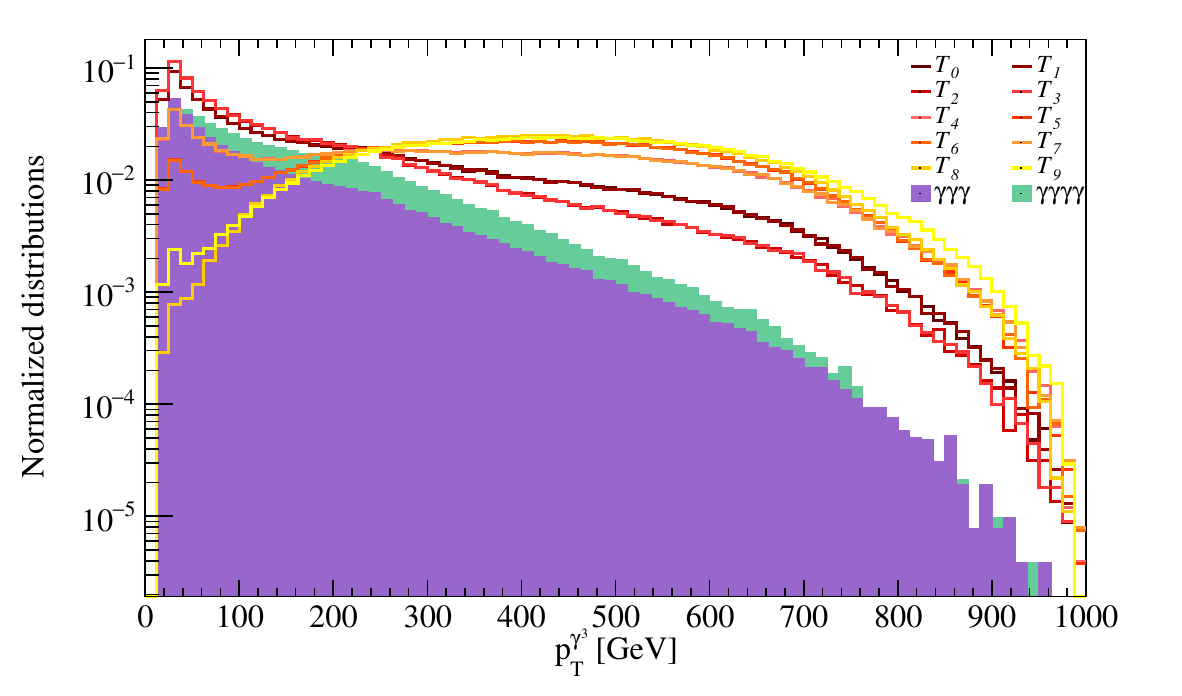}
\includegraphics[width=0.45\textwidth]{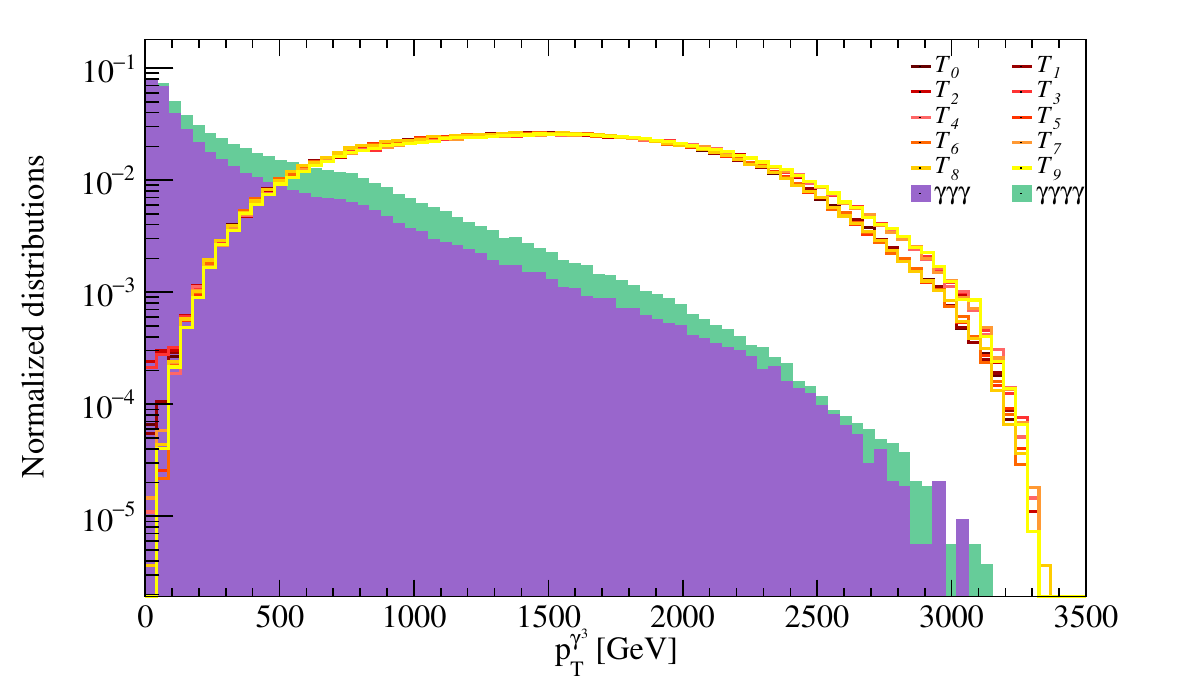}  \\
\caption{Normalized distributions of the third photon transverse momentum ($p_{T}^{\gamma_3}$) for the $\gamma\gamma\gamma$ final state at center-of-mass energies of $\sqrt{s} = 3~\mathrm{TeV}$ (left) and $10~\mathrm{TeV}$ (right).}
\label{fig:triphtonpt}
\end{figure*}

To select the signal events for this process, we first apply primary selection criteria requiring at least three reconstructed photons satisfying $p_{T}^{\gamma} > 20~\mathrm{GeV}$ and $|\eta_{\gamma}| < 2.5$. Furthermore, to ensure well-isolated photon candidates and avoid overlap, we enforce an angular separation requirement of $\Delta R_{\gamma\gamma} > 0.4$ between every pair of photons. To suppress the contributions of background processes, we examine various kinematic distributions to identify optimal secondary selection cuts. Among the observables evaluated, the transverse momentum of the third photon ($p_{T}^{\gamma_3}$) exhibits the strongest discrimination power between the signal and background events. The corresponding distributions for $\sqrt{s} = 3$ and $10~\mathrm{TeV}$ are displayed in Figure~\ref{fig:triphtonpt}.

The features of the photon transverse momentum distributions for the triphoton signal can be understood through the interplay between operator tensor structures and high-energy scaling behavior. Because all dimension-eight aQGC operators are constructed from covariant derivatives and field strength tensors, their amplitudes grow with momentum transfer, resulting in significantly harder $p_T$ spectra with broader tails compared to the steeply falling SM background. At $\sqrt{s} = 3\text{ TeV}$, subtle differences in Lorentz index contractions and derivative placements among individual $T_0$--$T_9$ operators, coupled with the linear interference term, lead to noticeable variations in shape and spread across the curves. However, as the collision energy increases to $10\text{ TeV}$, the cross-section parameterization (see Eq.~\ref{eq:total_xsection}) dictates that the pure quadratic BSM contribution scales aggressively as $\sigma_{\text{BSM}} \propto s^3 (f_i/\Lambda^4)^2$. This rapid power-law growth completely dominates the high-energy regime, washing out operator-specific variations and forcing all $p_T$ distributions to stretch out and converge into a uniform plateau.

To optimize the separation between signal events and the SM background, we implement secondary selection cuts based on the transverse momentum of the softest photon in the final state ($p_{T}^{\gamma_3}$). Specifically, we apply the following kinematic bounds:
\begin{align}
\label{eq:pt_cuts_3TeV}
200~\mathrm{GeV} < p_{T}^{\gamma_3} &< 1000~\mathrm{GeV} \quad (\text{for } \sqrt{s} = 3~\mathrm{TeV}), \\
\label{eq:pt_cuts_10TeV}
520~\mathrm{GeV} < p_{T}^{\gamma_3} &< 3500~\mathrm{GeV} \quad (\text{for } \sqrt{s} = 10~\mathrm{TeV}).
\end{align}

These cuts effectively suppress the background while focusing on the high-energy regions where the signal effects stand out most clearly. 
The cut-flow for cross~sections for the signal and SM background processes at both center-of-mass energies are summarized 
in Table~\ref{tab:threephotoncutflow}. These efficiency values are subsequently employed in the statistical 
analysis presented in Section~\ref{sec:result} to derive constraints on the Wilson coefficients.

\begin{table}[htbp]
\centering
\caption{Cut-flow table for cross~sections for the $\mu^+\mu^- \to \gamma\gamma\gamma$ signal (evaluated at a benchmark value of $f_{T_i}/\Lambda^4 = 10^{-12}\text{ GeV}^{-4}$) and corresponding SM background processes at center-of-mass energies of $\sqrt{s} = 3~\mathrm{TeV}$ and $10~\mathrm{TeV}$.}

\label{tab:threephotoncutflow}

\small
\renewcommand{\arraystretch}{1.3}
\setlength{\tabcolsep}{4pt}

\begin{tabular}{|c|ccc|ccc|}
\hline
&
\multicolumn{3}{c|}{$\sqrt{s}=3\,\mathrm{TeV}$}
&
\multicolumn{3}{c|}{$\sqrt{s}=10\,\mathrm{TeV}$} \\
\hline

Coupling
& \shortstack{\\Generation-\\Level Cuts\\(pb)}
& \shortstack{\\Primary\\Cuts\\(pb)}
& \shortstack{\\Secondary\\Cuts\\(pb)}
& \shortstack{\\Generation-\\Level Cuts\\(pb)}
& \shortstack{\\Primary\\Cuts\\(pb)}
& \shortstack{\\Secondary\\Cuts\\(pb)} \\
\hline

$f_{T_0}$
& $ 7.676 \times10^{-3} $ & $ 5.039 \times10^{-3} $  & $ 2.190 \times10^{-3} $ & 1.808 & 1.416 & 1.359 \\

$f_{T_1}$
& $ 7.676 \times10^{-3} $ & $ 5.034 \times10^{-3} $  & $ 2.184 \times10^{-3} $ & 1.808 & 1.414 & 1.357 \\

$f_{T_2}$
& $ 6.565 \times10^{-3} $ & $ 4.168 \times10^{-3} $ & $ 1.395 \times10^{-3} $ & 0.458 & 0.359 & 0.346 \\

$f_{T_3}$
& $ 6.565 \times10^{-3} $ & $ 4.173 \times10^{-3} $ & $ 1.397 \times10^{-3} $ & 0.458 & 0.360 & 0.347 \\

$f_{T_4}$
& $ 1.52 \times10^{-2} $ & $ 1.10 \times10^{-2} $ & $ 7.635 \times10^{-3} $ & 10.14 & 7.968 & 7.674 \\

$f_{T_5}$
& $ 3.773 \times10^{-2} $ & $ 2.867 \times10^{-2} $ & $ 2.383 \times10^{-2} $ & 40.26 & 31.537 & 30.301 \\

$f_{T_6}$
& $ 3.773 \times10^{-2} $ & $ 2.872 \times10^{-2} $ & $ 2.389 \times10^{-2} $ & 40.26 & 31.563 & 30.322 \\

$f_{T_7}$
& $ 1.52 \times10^{-2} $ & $ 1.10 \times10^{-2} $ & $ 7.639 \times10^{-3} $ & 10.14 & 7.972 & 7.679 \\

$f_{T_8}$
& 1.153 & $ 9.026 \times10^{-1} $ & $ 8.230 \times10^{-1} $ & 1551 & 1214.7 & 1167 \\

$f_{T_9}$
& $ 3.021 \times10^{-1} $ & $ 2.380 \times10^{-1} $ & $ 2.15 \times10^{-1} $ & 390.4 & 306.652 & 295.6 \\

\hline

$ \gamma\gamma\gamma\,(\mathrm{SM})$
& $ 5.364 \times10^{-3} $ & $ 3.313 \times10^{-3} $ & $ 8.608 \times10^{-4} $ & $ 6.341 \times10^{-4} $ & $ 4.158 \times10^{-4} $ & $ 1.030 \times10^{-4} $ \\

$ \gamma\gamma\gamma\gamma$
& $ 2.27 \times10^{-4} $ & $ 1.966 \times10^{-4} $ & $ 6.281 \times10^{-5} $ & $3.584\times10^{-5}$ & $ 3.198 \times10^{-5} $ & $ 1.025 \times10^{-5} $ \\

\hline
\end{tabular}
\end{table}

\subsubsection{The $\gamma\gamma Z$ Final State}

The second process analyzed in this study has a final state consisting of two photons and a $Z$ boson ($\gamma\gamma Z$). While its inclusive production cross~section exceeds that of the triphoton channel, requiring a leptonic decay of the $Z$ boson ($Z \to l^+l^-$ with $l = e, \mu$) results in a reduced visible rate for the $\gamma\gamma l^+l^-$ signature. Since all ten transverse operators $\mathcal{O}_{T_i}$ induce neutral quartic couplings, they all actively contribute to this final state. Following the same methodology, the signal cross~sections extracted from \textsc{MadGraph5\_aMC@NLO} simulations are evaluated at $\sqrt{s} = 3$ and $10~\mathrm{TeV}$ and parameterized as quadratic functions of the respective Wilson coefficients, as detailed in Table~\ref{tab:twophotonfit}.

The primary irreducible SM background originates from $\gamma \gamma l^+l^-$ production, alongside reducible contributions from $\gamma\gamma W^+W^-$ processes with subsequent leptonic decays of the $W$ bosons. Additional background contributions from $hZ \to \gamma\gamma l^+l^-$ and $\gamma\gamma\tau^+\tau^- \to \gamma\gamma l^+l^-\nu\bar{\nu}$ were also evaluated. However, owing to their smaller production cross~sections relative to the signal, these channels are found to be negligible and are thus omitted from the subsequent analysis.

\begin{table}[t]
\centering
\caption{Quadratic fits to the cross~sections as functions of the effective coupling parameter $\lambda = f_{T_i}/\Lambda^4$ for the $\gamma\gamma Z$ final state at center-of-mass energies of $\sqrt{s} = 3$ and $10~\mathrm{TeV}$.}

\scriptsize
\renewcommand{\arraystretch}{1.1}
\setlength{\tabcolsep}{3pt}

\begin{adjustbox}{max width=\textwidth}
\begin{tabular}{|c|c|c|}
\hline
Parameter $\lambda$ (GeV$^{-4}$) &
Quadratic functions for $\gamma\gamma Z$ final state at $3$ TeV (pb) &
Quadratic functions for $\gamma\gamma Z$ final state at $10$ TeV (pb) \\
\hline
$f_{T_0}/\Lambda^4 (f_{T_1}/\Lambda^4) $ &
$\sigma(\lambda)= 1.56  \times10^{-4} + 3.41 \times10^{7}\lambda + 3.07 \times10^{20}\lambda^{2} $ &
$\sigma(\lambda)= 6.24 \times10^{-6} + 3.66 \times10^{7}\lambda + 5.05 \times10^{22}\lambda^{2} $ \\
\hline
$f_{T_2}/\Lambda^4 (f_{T_3}/\Lambda^4) $ &
$\sigma(\lambda)= 1.56  \times10^{-4} + 2.15 \times10^{7}\lambda + 7.94 \times10^{19}\lambda^{2} $ &
$\sigma(\lambda)= 6.24 \times10^{-6} + 8.26 \times10^{6}\lambda + 1.21 \times10^{22}\lambda^{2} $ \\
\hline
$f_{T_4}/\Lambda^4$ &
$\sigma(\lambda)= 1.56  \times10^{-4} - 1.00 \times10^{6}\lambda + 2.36 \times10^{20}\lambda^{2} $ &
$\sigma(\lambda)= 6.24 \times10^{-6} + 1.40 \times10^{8}\lambda + 3.77 \times10^{22}\lambda^{2} $ \\
\hline
$f_{T_5}/\Lambda^4$ &
$\sigma(\lambda)= 1.56  \times10^{-4} + 8.27 \times10^{6}\lambda + 1.41 \times10^{21}\lambda^{2} $ &
$\sigma(\lambda)= 6.24 \times10^{-6} - 1.25 \times10^{9}\lambda+ 2.38 \times10^{23}\lambda^{2} $ \\
\hline
$f_{T_6}/\Lambda^4$ &
$\sigma(\lambda)= 1.56  \times10^{-4} + 1.14 \times10^{5}\lambda  + 8.63 \times10^{20}\lambda^{2} $ &
$\sigma(\lambda)= 6.24 \times10^{-6} - 4.27 \times10^{8}\lambda + 1.45 \times10^{23}\lambda^{2} $ \\
\hline
$f_{T_7}/\Lambda^4$ &
$\sigma(\lambda)= 1.56  \times10^{-4} +1.64 \times10^{6}\lambda  + 1.83 \times10^{20}\lambda^{2} $ &
$\sigma(\lambda)= 6.24 \times10^{-6} - 1.19 \times10^{8}\lambda + 3.01 \times10^{22}\lambda^{2} $ \\
\hline
$f_{T_8}/\Lambda^4$ &
$\sigma(\lambda)= 1.56  \times10^{-4} + 2.15 \times10^{8}\lambda + 2.45 \times10^{22}\lambda^{2} $ &
$\sigma(\lambda)= 6.24 \times10^{-6} - 1.02 \times10^{10}\lambda + 4.05 \times10^{24}\lambda^{2} $ \\
\hline
$f_{T_9}/\Lambda^4$ &
$\sigma(\lambda)= 1.56  \times10^{-4} + 1.06 \times10^{8}\lambda + 5.99 \times10^{21}\lambda^{2} $ &
$\sigma(\lambda)= 6.24 \times10^{-6} + 6.44 \times10^{8}\lambda + 9.63 \times10^{23}\lambda^{2} $ \\
\hline
\end{tabular}
\end{adjustbox}
\label{tab:twophotonfit}
\end{table}

Event selection for this channel proceeds through a set of primary selection criteria designed for the final-state signature, demanding at least two reconstructed photons with $p_{T}^{\gamma} > 20~\mathrm{GeV}$ and $|\eta_{\gamma}| < 2.5$, accompanied by at least two charged leptons satisfying $p_{T}^{l} > 20~\mathrm{GeV}$ and $|\eta_{l}| < 2.5$. Furthermore, to guarantee well-isolated candidates and avoid overlap, we enforce minimum angular separation bounds of $\Delta R_{\gamma l} > 0.4$. Afterward, the $Z$-boson candidate is reconstructed from opposite-sign same-flavor (OSSF) lepton pairs by requiring their invariant mass to fall within the window $76~\mathrm{GeV} < m_{\ell\ell} < 106~\mathrm{GeV}$.

\begin{figure*}[htbp]
\begin{center}   
        \includegraphics[width=0.45\textwidth]{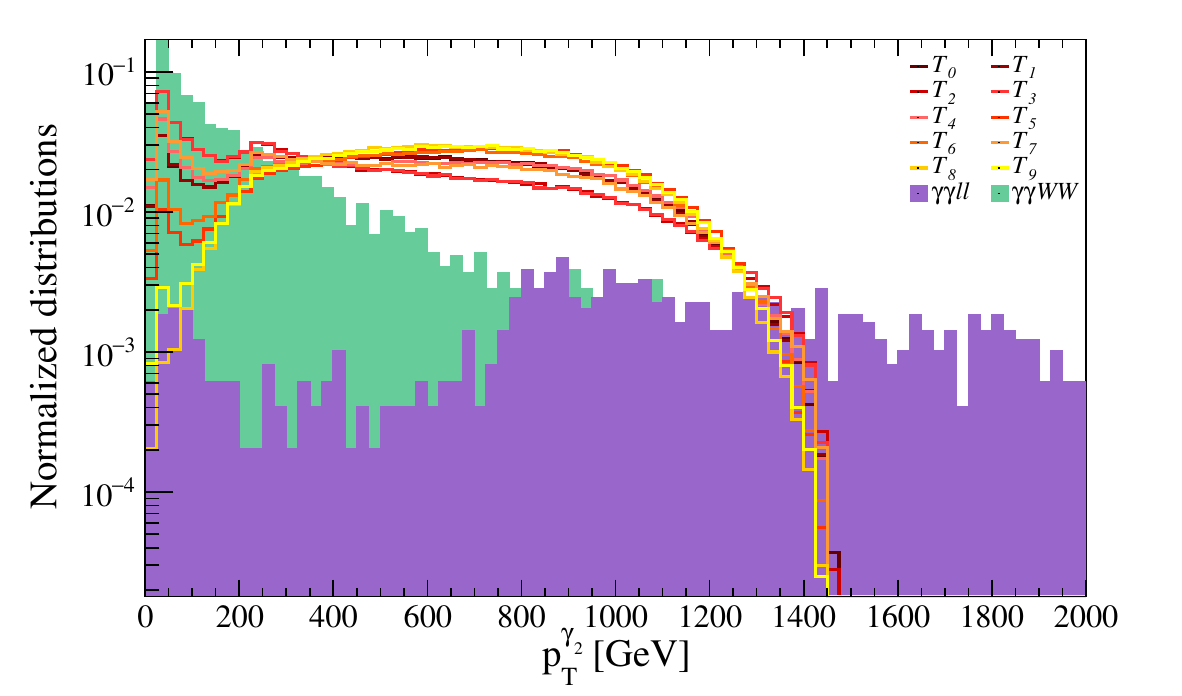}
        \includegraphics[width=0.45\textwidth]{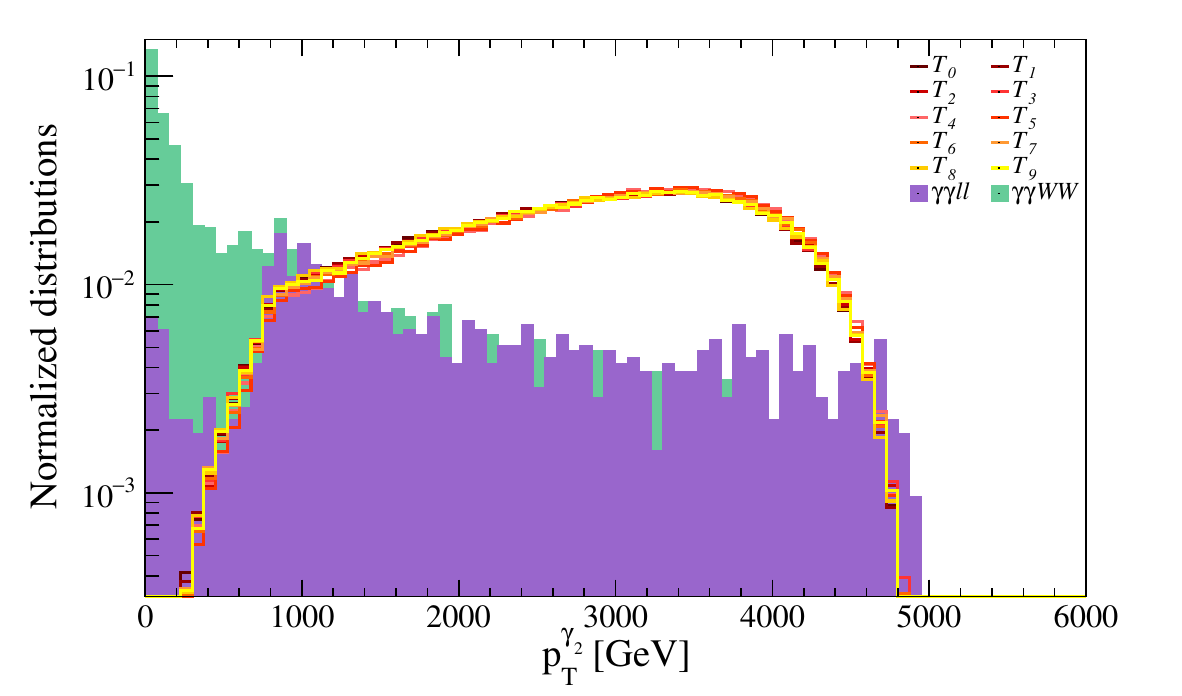}  \\
\caption{Normalized distributions of the second photon transverse momentum ($p_{T}^{\gamma_2}$) for the $\gamma\gamma Z$ final state at center-of-mass energies of $\sqrt{s} = 3~\mathrm{TeV}$ (left) and $10~\mathrm{TeV}$ (right).}
\label{fig:twophotonpt}
\end{center}
\end{figure*}

To further suppress background contributions, we explore various kinematic variables to determine effective secondary cuts. Among the tested observables, the transverse momentum of the second photon ($p_{T}^{\gamma_2}$) has strong separation power. Figure~\ref{fig:twophotonpt} displays these key distributions at $\sqrt{s} = 3$ and $10~\mathrm{TeV}$. The characteristic behavior of these transverse momentum distributions can be understood through the same operator tensor structures and high-energy scaling mechanisms discussed for the triphoton final state.

To further refine signal isolation and suppress background contributions, we apply a secondary selection cut focused on the transverse momentum of the second photon ($p_{T}^{\gamma_2}$). Specifically, we impose the following kinematic windows:
\begin{align}
250~\mathrm{GeV} < p_{T}^{\gamma_2} &< 1400~\mathrm{GeV} \quad (\text{for } \sqrt{s} = 3~\mathrm{TeV}), \label{eq:ggz_pt_cuts_3TeV} \\
1200~\mathrm{GeV} < p_{T}^{\gamma_2} &< 4600~\mathrm{GeV} \quad (\text{for } \sqrt{s} = 10~\mathrm{TeV}). \label{eq:ggz_pt_cuts_10TeV}
\end{align}

By focusing on these regions, these cuts effectively reduce the background while keeping the area where signal effects are strongest. The resulting cut-flow for cross~sections for both the signal benchmarks and the relevant SM backgrounds are detailed in Table~\ref{tab:twophotoncutflow}. These values serve as foundational inputs for the statistical evaluation in Section~\ref{sec:result}, where exclusion limits on the Wilson coefficients are established.

\begin{table}[htbp]
\centering
\caption{Cut-flow table for cross~sections for the $\mu^+\mu^- \to \gamma\gamma Z$ signal (evaluated at a benchmark value of $f_{T_i}/\Lambda^4 = 10^{-12}\text{ GeV}^{-4}$) and corresponding SM background processes at center-of-mass energies of $\sqrt{s} = 3~\mathrm{TeV}$ and $10~\mathrm{TeV}$.}
\label{tab:twophotoncutflow}
\small
\renewcommand{\arraystretch}{1.3}
\setlength{\tabcolsep}{4pt}

\begin{tabular}{|c|ccc|ccc|}
\hline
&
\multicolumn{3}{c|}{$\sqrt{s}=3\,\mathrm{TeV}$}
&
\multicolumn{3}{c|}{$\sqrt{s}=10\,\mathrm{TeV}$} \\
\hline

Coupling
& \shortstack{\\Generation-\\Level Cuts\\(pb)}
& \shortstack{\\Primary\\Cuts\\(pb)}
& \shortstack{\\Secondary\\Cuts\\(pb)}
& \shortstack{\\Generation-\\Level Cuts\\(pb)}
& \shortstack{\\Primary\\Cuts\\(pb)}
& \shortstack{\\Secondary\\Cuts\\(pb)} \\
\hline

$f_{T_0}$
& $ 4.98 \times10^{-4}$ &  $ 2.683 \times10^{-4}$ &  $ 2.158 \times10^{-4}$ & $ 5.05 \times10^{-2}$ & $ 2.18 \times10^{-2}$ & $ 2.00 \times10^{-2}$ \\

$f_{T_1}$
& $ 4.98 \times10^{-4}$ &  $ 2.680 \times10^{-4}$ &  $ 2.156 \times10^{-4}$ & $ 5.05 \times10^{-2}$ & $ 2.19 \times10^{-2}$ & $ 2.00 \times10^{-2}$ \\

$f_{T_2}$
& $ 2.568 \times10^{-4}$ &  $ 1.304 \times10^{-4}$ &  $ 8.701 \times10^{-5}$ & $ 1.213 \times10^{-2}$ & $ 5.186 \times10^{-3}$ & $ 4.756 \times10^{-3}$ \\

$f_{T_3}$
& $ 2.568 \times10^{-4}$ &  $ 1.301 \times10^{-4}$ &  $ 8.683 \times10^{-5}$ & $ 1.213 \times10^{-2}$ & $ 5.185 \times10^{-3}$ & $ 4.756 \times10^{-3}$ \\

$f_{T_4}$
& $ 2.924 \times10^{-4}$ &  $ 1.550 \times10^{-4}$ &  $ 1.203 \times10^{-4}$ & $ 3.794 \times10^{-2}$ & $ 1.63 \times10^{-2}$ & $ 1.50 \times10^{-2}$ \\

$f_{T_5}$
& $ 1.578 \times10^{-3}$ &  $ 8.945 \times10^{-4}$ &  $ 8.113 \times10^{-4}$ & $ 2.363 \times10^{-1}$ & $ 1.031 \times10^{-1}$ & $ 9.55 \times10^{-2}$ \\

$f_{T_6}$
& $ 1.027 \times10^{-3}$ &  $ 5.737 \times10^{-4}$ &  $ 5.042 \times10^{-4}$ & $ 1.446 \times10^{-1}$ & $ 6.25 \times10^{-2}$ & $ 5.76 \times10^{-2}$ \\

$f_{T_7}$
& $ 3.43 \times10^{-4}$ &  $ 1.799 \times10^{-4}$ &  $ 1.345 \times10^{-4}$ & $ 2.995 \times10^{-2}$ & $ 1.28 \times10^{-2}$ & $ 1.18 \times10^{-2}$ \\

$f_{T_8}$
& $ 2.49 \times10^{-2}$ & $ 1.42 \times10^{-2}$ & $ 1.33 \times10^{-2}$ & 4.043 & 1.743 & 1.595 \\

$f_{T_9}$
& $ 6.263 \times10^{-3}$ & $ 3.577 \times10^{-3}$ & $ 3.317 \times10^{-3}$ & $ 9.662 \times10^{-1}$ & $ 4.13 \times10^{-1}$ & $ 3.79 \times10^{-1}$ \\

\hline
$ \gamma \gamma l^{+} l^{-} (\mathrm{SM})$
& $ 5.029 \times10^{-3}$ & $ 1.577 \times10^{-5}$ & $ 5.431 \times10^{-6}$ & $ 7.7 \times10^{-4}$ & $ 2.415 \times10^{-6}$ & $ 1.558 \times10^{-6}$ \\

$ \gamma \gamma W^{+} W^{-} $
& $6.749\times10^{-5}$ & $ 8.365 \times10^{-7}$ & $ 2.439 \times10^{-7}$ & $1.491 \times10^{-5}$ & $ 8.678 \times10^{-8}$ & $ 2.441 \times10^{-8}$ \\

\hline
\end{tabular}
\end{table}

\subsubsection{The $\gamma ZZ$ Final State}

The third process under consideration involves a single photon produced in association with two $Z$ bosons ($\gamma ZZ$). Requiring both $Z$ bosons to undergo leptonic decays ($\gamma ZZ \to \gamma l^+l^-l^+l^-$ with $l = e, \mu$) imposes a substantial branching-ratio penalty, resulting in a considerably smaller visible cross~section compared to the previous channels. Consequently, this topology yields lower event counts and a correspondingly reduced statistical sensitivity to the target Wilson coefficients.
Because all ten transverse operators $\mathcal{O}_{T_i}$ give rise to neutral aQGC, each contributes to this final state. Following a parallel procedure, we generate signal events using \textsc{MadGraph5\_aMC@NLO} at center-of-mass energies of $\sqrt{s} = 3$ and $10~\mathrm{TeV}$. The resulting cross~sections are then modeled as quadratic functions of the individual Wilson coefficients, as listed in Table~\ref{tab:onephotonfit}.

The main irreducible SM background stems from $\gamma l^+l^-l^+l^-$ production, supplemented by reducible $Z W^+W^-$ processes with subsequent leptonic decays of the $W$ bosons. Additional processes---including $ZZ \to l^+l^-l^+l^-$, $ZZZ \to \nu\bar{\nu}l^+l^-l^+l^-$, and $h\gamma \to \gamma ZZ \to \gamma l^+l^-l^+l^-$---were also evaluated. However, due to their exceptionally low production rates compared to the signal and/or strong suppression from the applied selection cuts, these channels are safely neglected in the subsequent analysis.

\begin{table}[ht!]
\centering
\caption{Quadratic fits to the cross~sections as functions of the effective coupling parameter $\lambda = f_{T_i}/\Lambda^4$ for the $\gamma ZZ$ final state at center-of-mass energies of $\sqrt{s} = 3$ and $10~\mathrm{TeV}$.}
\scriptsize
\renewcommand{\arraystretch}{1.1}
\setlength{\tabcolsep}{3pt}

\begin{adjustbox}{max width=\textwidth}
\begin{tabular}{|c|c|c|}
\hline
Parameter $\lambda$ (GeV$^{-4}$) &
Quadratic functions for $\gamma Z Z$ final state at $3$ TeV (pb) &
Quadratic functions for $\gamma Z Z$ final state at $10$ TeV (pb) \\
\hline
$f_{T_0}/\Lambda^4 (f_{T_1}/\Lambda^4) $ &
$\sigma(\lambda)= 1.63 \times10^{-6} + 9.71 \times10^{5}\lambda + 2.13 \times10^{19}\lambda^{2} $ &
$\sigma(\lambda)= 8.78 \times10^{-9} - 1.48 \times10^{6}\lambda + 2.53 \times10^{20}\lambda^{2} $ \\
\hline
$f_{T_2}/\Lambda^4 (f_{T_3}/\Lambda^4) $ &
$\sigma(\lambda)= 1.63 \times10^{-6} + 7.64 \times10^{5}\lambda + 5.05 \times10^{18}\lambda^{2} $ &
$\sigma(\lambda)= 8.78 \times10^{-9} - 2.40 \times10^{5}\lambda + 5.92 \times10^{19}\lambda^{2} $ \\
\hline
$f_{T_4}/\Lambda^4$ &
$\sigma(\lambda)= 1.63 \times10^{-6} - 1.53 \times10^{5}\lambda + 4.28 \times10^{18}\lambda^{2} $ &
$\sigma(\lambda)= 8.78 \times10^{-9} - 2.33 \times10^{5}\lambda + 4.83 \times10^{19}\lambda^{2} $ \\
\hline
$f_{T_5}/\Lambda^4$ &
$\sigma(\lambda)= 1.63 \times10^{-6} - 9.82 \times10^{5} \lambda + 3.92 \times10^{19}\lambda^{2}  $ &
$\sigma(\lambda)= 8.78 \times10^{-9} - 1.72 \times10^{6}\lambda + 4.46 \times10^{20}\lambda^{2} $ \\
\hline
$f_{T_6}/\Lambda^4$ &
$\sigma(\lambda)= 1.63 \times10^{-6} - 1.03 \times10^{5}\lambda + 1.31 \times10^{19}\lambda^{2}  $ &
$\sigma(\lambda)= 8.78 \times10^{-9} - 1.01 \times10^{6}\lambda + 1.47 \times10^{20}\lambda^{2} $ \\
\hline
$f_{T_7}/\Lambda^4$ &
$\sigma(\lambda)= 1.63 \times10^{-6} - 2.00 \times10^{5}\lambda + 2.15 \times10^{18}\lambda^{2} $ &
$\sigma(\lambda)= 8.78 \times10^{-9} - 1.51 \times10^{5}\lambda  + 2.44 \times10^{19}\lambda^{2} $ \\
\hline
$f_{T_8}/\Lambda^4$ &
$\sigma(\lambda)= 1.63 \times10^{-6} + 1.52 \times10^{6}\lambda + 1.59 \times10^{20}\lambda^{2} $ &
$\sigma(\lambda)= 8.78 \times10^{-9} - 4.12 \times10^{6}\lambda + 1.88 \times10^{21}\lambda^{2} $ \\
\hline
$f_{T_9}/\Lambda^4$ &
$\sigma(\lambda)= 1.63 \times10^{-6} + 1.88 \times10^{6}\lambda + 3.80 \times10^{19}\lambda^{2} $ &
$\sigma(\lambda)= 8.78 \times10^{-9} - 2.66 \times10^{6}\lambda + 4.41 \times10^{20}\lambda^{2} $ \\
\hline
\end{tabular}
\end{adjustbox}
\label{tab:onephotonfit}
\end{table}

Event selection for this topology begins by requiring at least one photon with $p_{T}^{\gamma} > 20~\mathrm{GeV}$ and $|\eta_{\gamma}| < 2.5$, accompanied by at least four charged leptons satisfying $p_{T}^{l} > 20~\mathrm{GeV}$ and $|\eta_{l}| < 2.5$. To ensure reliable object identification and prevent detector overlap, we enforce a minimum angular separation of $\Delta R_{\gamma l} > 0.4$ between all photon-lepton pairs. 

Subsequently, the two $Z$ bosons are reconstructed from OSSF lepton pairs. Depending on the flavor composition of the four-lepton system, the events fall into three distinct categories:
\begin{enumerate}
    \item At least four electrons with no muons ($4e0\mu$),
    \item At least four muons with no electrons ($0e4\mu$),
    \item At least two electrons and two muons ($2e2\mu$).
\end{enumerate}
Within each category, OSSF lepton pairs are systematically paired to identify the combination whose invariant mass lies closest to the nominal $Z$-boson mass. The primary $Z$-boson candidate is defined as the pair whose invariant mass falls closest to the $Z$ pole and satisfies the mass window $76~\mathrm{GeV} < m_{\ell\ell} < 106~\mathrm{GeV}$.

\begin{figure*}[htbp]
\begin{center}
        \includegraphics[width=0.45\textwidth]{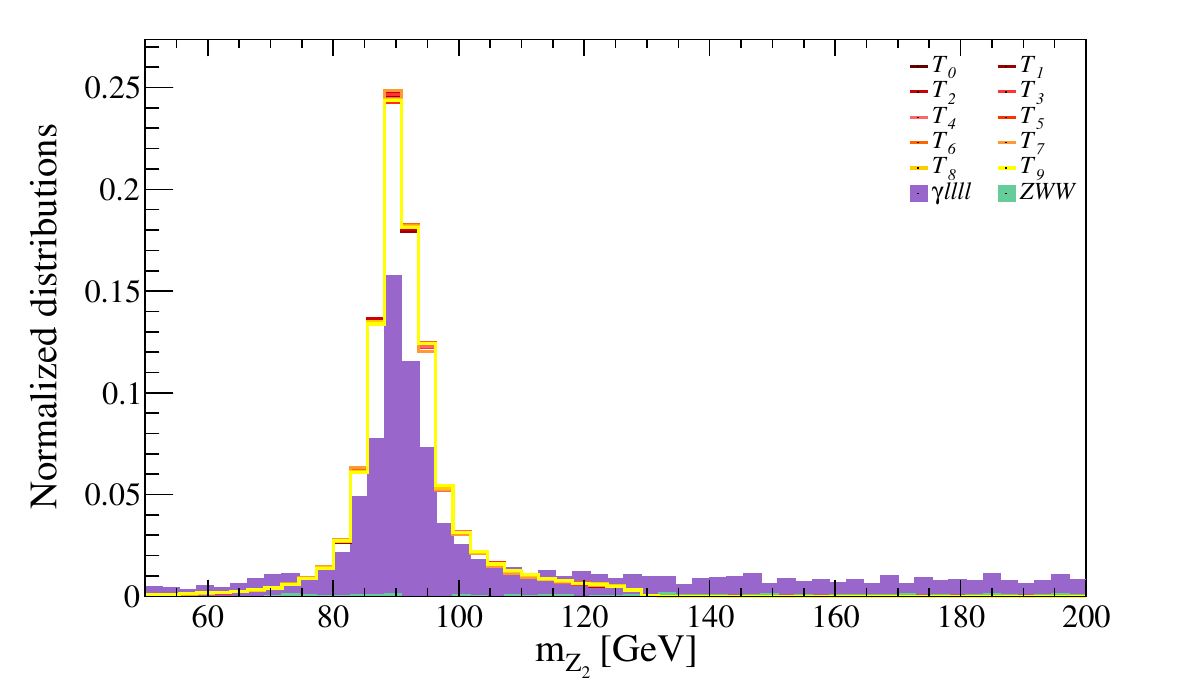}  
        \includegraphics[width=0.45\textwidth]{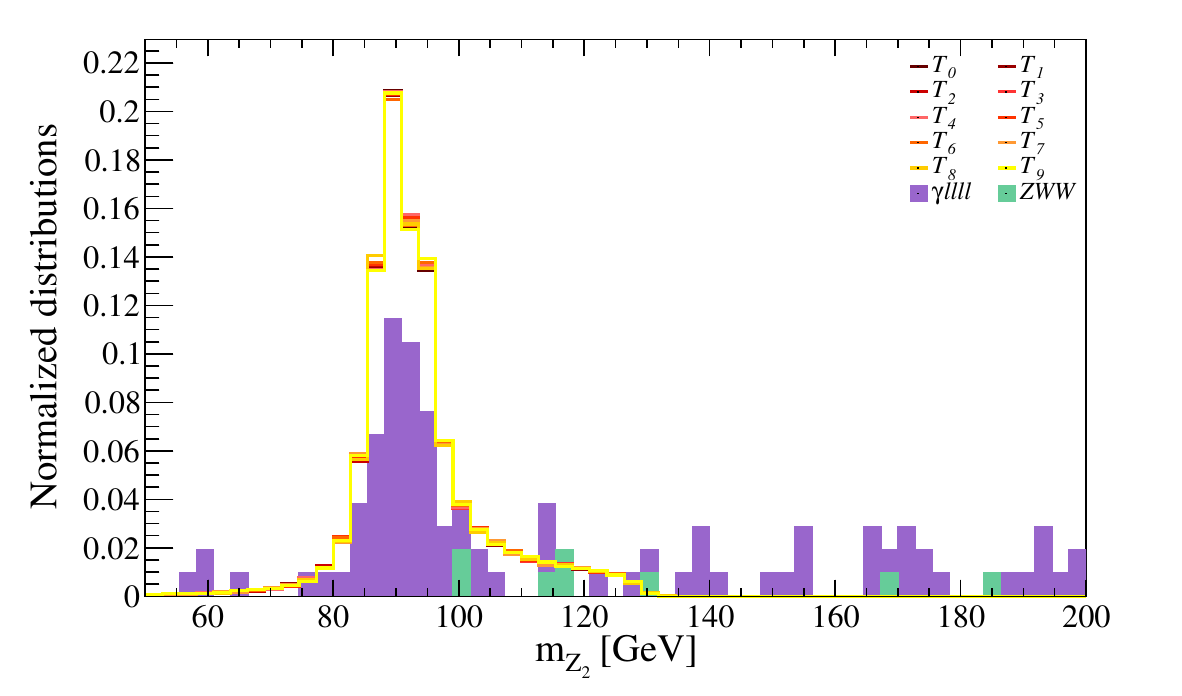}  \\
\caption{Normalized distributions of the invariant mass of the secondary reconstructed $Z$ boson for the $\gamma ZZ$ final state at center-of-mass energies of $\sqrt{s} = 3~\mathrm{TeV}$ (left) and $10~\mathrm{TeV}$ (right).}
		\label{fig:threephotonpt}
	\end{center}
\end{figure*}

To further reduce the background, we examine other kinematic variables to set effective secondary cuts. 
Among the evaluated observables, the invariant mass of the secondary $Z$ boson ($m_{Z_2}$) exhibits strong discrimination power; inspection of its distribution reveals that the background forms a relatively broad, nearly uniform spectrum, whereas the signal features a pronounced peak near the nominal $Z$-boson mass ($m_Z \simeq 90~\mathrm{GeV}$). Figure~\ref{fig:threephotonpt} presents these differential spectra at $\sqrt{s} = 3$ and $10~\mathrm{TeV}$.

To maximize signal purity against SM backgrounds, we apply a secondary mass window requirement on the second $Z$-boson candidate ($m_{Z_2}$). Specifically, we enforce the following kinematic bound:
\begin{align}
76~\mathrm{GeV} < m_{Z_2} < 106~\mathrm{GeV} \quad (\text{for both } \sqrt{s} = 3 \text{ and } 10~\mathrm{TeV}).
\end{align}

This condition efficiently suppresses background contributions while isolating the resonant signal region. The resulting cut-flow for cross~sections for both the signal benchmarks and the relevant background processes at both center-of-mass energies are compiled in Table~\ref{tab:onephotoncutflow}. Finally, these efficiencies feed directly into the statistical framework described in Section~\ref{sec:result} to set exclusion bounds on the Wilson coefficients.

\begin{table}[htbp]
\centering
\caption{Cut-flow table for cross~sections for the $\mu^+\mu^- \to \gamma Z Z$ signal (evaluated at a benchmark value of $f_{T_i}/\Lambda^4 = 10^{-12}\text{ GeV}^{-4}$) and corresponding SM background processes at center-of-mass energies of $\sqrt{s} = 3~\mathrm{TeV}$ and $10~\mathrm{TeV}$.}
\label{tab:onephotoncutflow}

\small
\renewcommand{\arraystretch}{1.3}
\setlength{\tabcolsep}{4pt}

\begin{tabular}{|c|ccc|ccc|}
\hline
&
\multicolumn{3}{c|}{$\sqrt{s}=3\,\mathrm{TeV}$}
&
\multicolumn{3}{c|}{$\sqrt{s}=10\,\mathrm{TeV}$} \\
\hline

Coupling
& \shortstack{\\Generation-\\Level Cuts\\(pb)}
& \shortstack{\\Primary\\Cuts\\(pb)}
& \shortstack{\\Secondary\\Cuts\\(pb)}
& \shortstack{\\Generation-\\Level Cuts\\(pb)}
& \shortstack{\\Primary\\Cuts\\(pb)}
& \shortstack{\\Secondary\\Cuts\\(pb)} \\
\hline

$f_{T_0}$
& $ 2.386 \times 10^{-5} $ & $ 1.056 \times 10^{-5} $ & $ 9.535 \times 10^{-6} $ & $ 2.52 \times 10^{-4} $ & $ 4.036 \times 10^{-5} $ & $ 3.507 \times 10^{-5} $ \\

$f_{T_1}$
& $ 2.386 \times 10^{-5} $ & $ 1.057 \times 10^{-5} $ & $ 9.546 \times 10^{-6} $ & $ 2.52 \times 10^{-4} $ & $ 4.062 \times 10^{-5} $ & $ 3.534 \times 10^{-5} $ \\

$f_{T_2}$
& $ 7.444 \times 10^{-6} $ & $ 3.222 \times 10^{-6} $ & $ 2.920 \times 10^{-6} $ & $ 5.891 \times 10^{-5} $  & $ 9.284 \times 10^{-6} $ & $ 8.055 \times 10^{-6} $ \\

$f_{T_3}$
& $ 7.444 \times 10^{-6} $ & $ 3.204 \times 10^{-6} $ & $ 2.904 \times 10^{-6} $ & $ 5.891 \times 10^{-5} $  & $ 9.349 \times 10^{-6} $  & $ 8.118 \times 10^{-6} $ \\

$f_{T_4}$
& $5.775 \times 10^{-6}$ & $ 2.473 \times 10^{-6}$ & $ 2.232 \times 10^{-6}$ & $4.814 \times 10^{-5}$ & $ 7.893 \times 10^{-6}$ & $ 6.848 \times 10^{-6}$ \\

$f_{T_5}$
& $3.993 \times 10^{-5}$ & $ 1.772 \times 10^{-5}$ & $ 1.600 \times 10^{-5}$ & $ 4.441 \times 10^{-4} $ & $ 7.178 \times 10^{-5} $ & $ 6.254 \times 10^{-5} $ \\

$f_{T_6}$
& $ 1.47 \times 10^{-5} $ &  $ 6.436 \times 10^{-6} $ & $ 5.816 \times 10^{-6} $ & $ 1.467 \times 10^{-4} $ & $ 2.333 \times 10^{-5} $ & $ 2.025 \times 10^{-5} $ \\

$f_{T_7}$
& $ 3.582 \times 10^{-6} $ &  $ 1.481 \times 10^{-6} $  &  $ 1.342 \times 10^{-6} $  & $ 2.432 \times 10^{-5} $ & $ 3.930 \times 10^{-6} $ & $ 3.414 \times 10^{-6} $ \\

$f_{T_8}$
& $ 1.626 \times 10^{-4} $ & $ 7.298 \times 10^{-5} $ & $ 6.598 \times 10^{-5} $ & $ 1.875 \times 10^{-3} $ & $ 2.987 \times 10^{-4} $ & $ 2.595 \times 10^{-4} $ \\

$f_{T_9}$
& $ 4.17 \times 10^{-5} $ & $ 1.863 \times 10^{-5} $ & $ 1.681 \times 10^{-5} $ & $ 4.38 \times 10^{-4} $ & $ 6.957 \times 10^{-5} $ & $ 6.032 \times 10^{-5} $ \\

\hline
$ \gamma l^{+} l^{-} l^{+} l^{-} (\mathrm{SM})$
& $ 1.094 \times 10^{-4} $ & $ 1.665 \times 10^{-5} $  & $ 6.764 \times 10^{-7} $  & $ 1.609 \times 10^{-5} $ & $ 1.692 \times 10^{-6} $ & $ 2.436 \times 10^{-9} $ \\

$ Z {W}^{+} {W}^{-} $
& $ 2.857 \times 10^{-5} $ & $ 2.307 \times 10^{-7} $ & $ 1.224 \times 10^{-9} $ & $ 2.618 \times 10^{-6} $ & $ 2.507 \times 10^{-8} $ & $ 1.496 \times 10^{-11} $ \\

\hline
\end{tabular}
\end{table}

\subsubsection{The $ZZZ$ Final State}

Finally, the fourth channel evaluates the production of three $Z$ bosons ($ZZZ$). To the best of our knowledge, this topology has not been studied before at any proposed collider. For simplicity, we focus on decays into electron-positron pairs. Although the multiple $Z$ decays greatly reduce the event rate and lead to weaker limits on the couplings, we include this channel to ensure a complete study of all neutral quartic interactions.
Since all ten transverse operators $\mathcal{O}_{T_i}$ generate neutral anomalous quartic couplings, every operator contributes to this reaction. Using the same workflow, we simulate signal samples with \textsc{MadGraph5\_aMC@NLO} at collision energy of $3\text{ TeV}$. We extract the quadratic cross~section parameterizations as summarized in Table~\ref{tab:nophotonfit}. The primary irreducible background arises from SM $e^+e^-e^+e^-e^+e^-$ production. 

\begin{table}[ht!]
\centering
\caption{Fits to the cross~sections as functions of the effective coupling parameter $\lambda = f_{T_i}/\Lambda^4$ for the $ZZZ$ final state at center-of-mass energies of $\sqrt{s} = 3$.}
\label{tab:nophotonfit}
\scriptsize
\renewcommand{\arraystretch}{1.1}
\setlength{\tabcolsep}{3pt}

\begin{adjustbox}{max width=\textwidth}
\begin{tabular}{|c|c|c|}
\hline
Parameter $\lambda$ (GeV$^{-4}$) &
Quadratic functions for $Z Z Z$ final state at $3$ TeV (pb) \\
\hline
$f_{T_0}/\Lambda^4 (f_{T_1}/\Lambda^4) $ &
$\sigma(\lambda)= 7.66 \times 10^{-10} + 1.71 \times 10^{3} \lambda + 7.08 \times 10^{16} \lambda^{2} $ \\
\hline
$f_{T_2}/\Lambda^4 (f_{T_3}/\Lambda^4) $ &
$\sigma(\lambda)= 7.66 \times 10^{-10} + 8.81 \times 10^{2} \lambda + 1.69 \times 10^{16} \lambda^{2} $ \\
\hline
$f_{T_4}/\Lambda^4 (f_{T_7}/\Lambda^4) $ &
$\sigma(\lambda)= 7.66 \times 10^{-10} + 4.16 \times 10^{2} \lambda + 6.27 \times 10^{15} \lambda^{2} $ \\
\hline
$f_{T_5}/\Lambda^4 (f_{T_6}/\Lambda^4) $ &
$\sigma(\lambda)= 7.66 \times 10^{-10} + 6.33 \times 10^{2} \lambda + 2.64 \times 10^{16} \lambda^{2} $ \\
\hline
$f_{T_8}/\Lambda^4$ &
$\sigma(\lambda)= 7.66 \times 10^{-10} + 22.7 \ \lambda + 4.93 \times 10^{16} \lambda^{2} $ \\
\hline
$f_{T_9}/\Lambda^4$ &
$\sigma(\lambda)= 7.66 \times 10^{-10} - 6.22 \ \lambda + 1.17 \times 10^{16} \lambda^{2}  $ \\
\hline
\end{tabular}
\end{adjustbox}
\end{table}

Event selection for this final state begins by requiring at least six electrons, each satisfying $p_{T}^{e} > 20~\mathrm{GeV}$ and $|\eta_e| < 2.5$. To reconstruct the three $Z$ bosons, we group the leptons into opposite-sign (OS) pairs. All possible OS combinations are evaluated to identify the triplet of pairs that best matches the nominal $Z$-boson mass. The resulting $Z$ candidates are then ordered by how closely their invariant masses approach the $Z$ pole. The primary $Z$ candidate is defined as the pair closest to the pole mass, which must fall within the mass window $76~\mathrm{GeV} < m_{\ell\ell} < 106~\mathrm{GeV}$.

\begin{figure*}[htbp]
\begin{center}
		\includegraphics[width=0.45\textwidth]{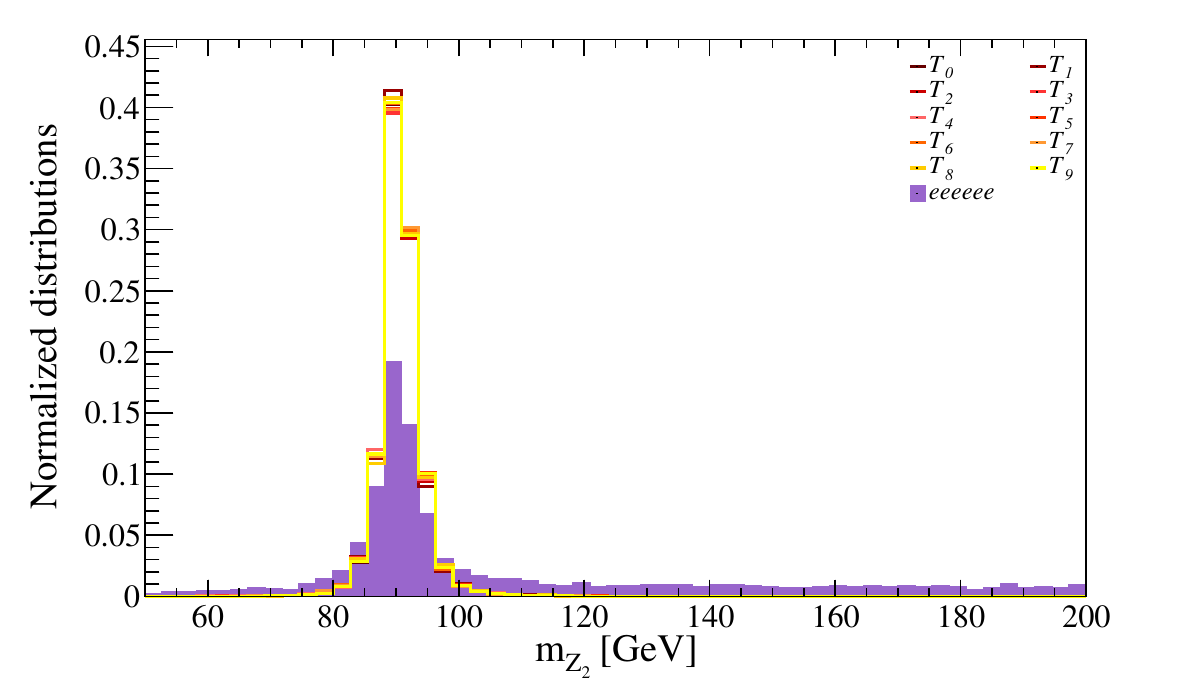}
		\includegraphics[width=0.45\textwidth]{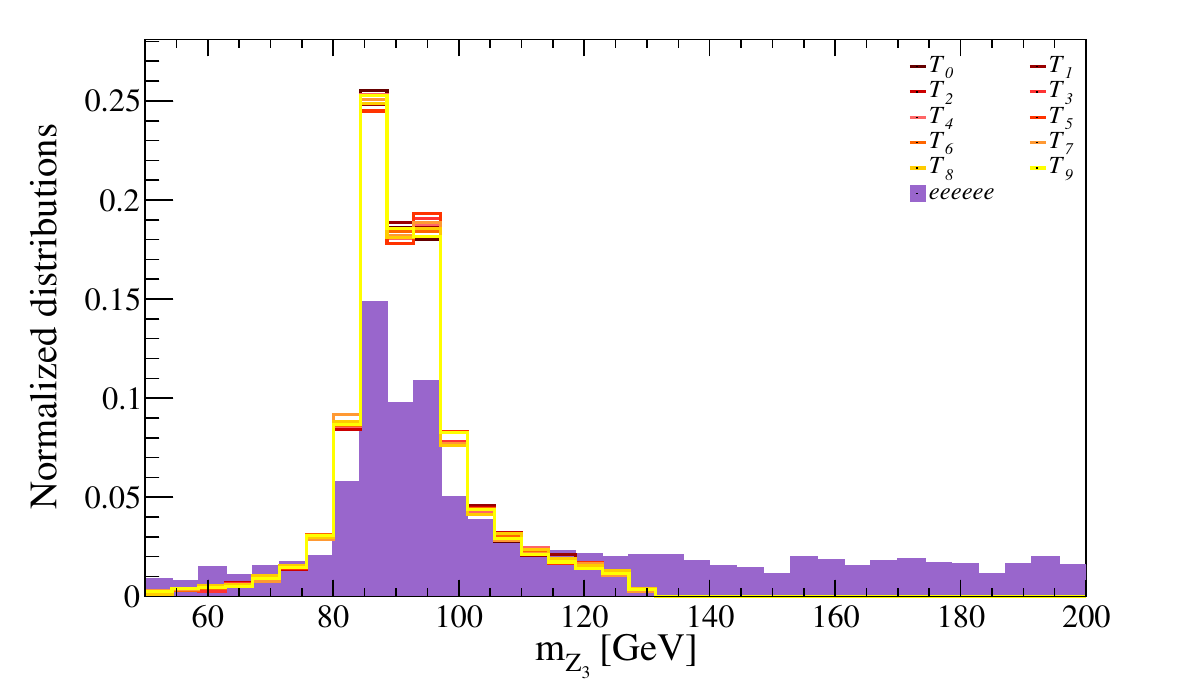}  \\
		\caption{Normalized distributions of the invariant masses of the secondary (left) and tertiary (right) reconstructed $Z$ bosons for the $ZZZ$ final state at center-of-mass energy of $\sqrt{s} = 3~\mathrm{TeV}$.}
		\label{processe4}
	\end{center}
\end{figure*}

To achieve additional background suppression, we turn to further kinematic properties of the multi-electron system. 
Specifically, the invariant masses of both the secondary ($m_{Z_2}$) and tertiary ($m_{Z_3}$) $Z$ bosons provide excellent separation power; their respective distributions show that background events are spread out nearly uniformly, whereas signal events display sharp peaks centered on the nominal $Z$ pole ($m_Z \simeq 90~\mathrm{GeV}$). Figure~\ref{processe4} illustrates these differential spectra at $\sqrt{s} = 3\text{ TeV}$.

To isolate the signal from remaining backgrounds, we apply mass window selections on both the secondary and tertiary $Z$ candidates. Explicitly, we impose the following kinematic bounds:
\begin{align}
76~\mathrm{GeV} < m_{Z_2} < 106~\mathrm{GeV} \quad \text{and} \quad 76~\mathrm{GeV} < m_{Z_3} < 106~\mathrm{GeV}.
\label{eq:zzz_mass_cuts}
\end{align}

These criteria successfully suppress non-resonant background contributions while retaining the signal acceptance. Table~\ref{tab:nophotoncutflow} lists the resulting cut-flow for cross~sections for both the benchmark signals and background processes. Ultimately, these efficiencies are incorporated into the statistical evaluation outlined in Section~\ref{sec:result} to extract exclusion limits on the Wilson coefficients.

 \begin{table}[htbp]
\centering
\caption{Cut-flow table for cross~sections for the $\mu^+\mu^- \to Z Z Z$ signal (evaluated at a benchmark value of $f_{T_i}/\Lambda^4 = 10^{-12}\text{ GeV}^{-4}$) and corresponding SM background processes at center-of-mass energy of $\sqrt{s} = 3~\mathrm{TeV}$ .}
\label{tab:nophotoncutflow}

\small
\renewcommand{\arraystretch}{1.3}
\setlength{\tabcolsep}{4pt}

\begin{tabular}{|c|ccc|}
\hline
&
\multicolumn{3}{c|}{$\sqrt{s}=3\,\mathrm{TeV}$} \\
\hline

Coupling
& \shortstack{\\Generation-\\Level Cuts\\(pb)}
& \shortstack{\\Primary\\Cuts\\(pb)}
& \shortstack{\\Secondary\\Cuts\\(pb)} \\
\hline

$f_{T_0}$
& $ 7.334 \times 10^{-8} $ & $ 1.470 \times 10^{-8} $ & $ 1.257 \times 10^{-8} $  \\

$f_{T_1}$
& $ 7.334 \times 10^{-8} $ & $ 1.482 \times 10^{-8} $ & $ 1.268 \times 10^{-8} $  \\

$f_{T_2}$
& $ 1.86 \times 10^{-8} $ & $ 3.696 \times 10^{-9} $ & $ 3.157 \times 10^{-9} $  \\

$f_{T_3}$
& $ 1.86 \times 10^{-8} $ & $ 3.639 \times 10^{-9} $ & $ 3.095 \times 10^{-9} $  \\

$f_{T_4}$
& $ 7.453 \times 10^{-9} $ & $ 1.454 \times 10^{-9} $ & $ 1.239 \times 10^{-9} $  \\

$f_{T_5}$
& $ 2.783 \times 10^{-8} $ & $ 5.566 \times 10^{-9} $  & $ 4.765 \times 10^{-9} $  \\

$f_{T_6}$
& $ 2.783 \times 10^{-8} $ & $ 5.575 \times 10^{-9} $  & $ 4.752 \times 10^{-9} $  \\

$f_{T_7}$
& $ 7.453 \times 10^{-9} $ & $ 1.453 \times 10^{-9} $  & $ 1.244 \times 10^{-9} $  \\

$f_{T_8}$
& $ 5.014 \times 10^{-8} $ & $ 1.021 \times 10^{-8} $  & $ 8.642 \times 10^{-9} $  \\

$f_{T_9}$
& $ 1.25 \times 10^{-8} $ & $ 2.473 \times 10^{-9} $  & $ 2.124 \times 10^{-9} $   \\

\hline
$ e^{+} e^{-} e^{+} e^{-} e^{+} e^{-} (\mathrm{SM})$
& $ 6.559 \times 10^{-9} $  & $ 9.600 \times 10^{-10} $  & $ 1.342 \times 10^{-10} $ \\

\hline
\end{tabular}
\end{table}

	\section{Results and Sensitivity Estimation} \label{sec:result}                                       

	In this section, we evaluate the projected sensitivity of a future high-energy muon collider to aQGCs through three neutral electroweak gauge boson production. Specifically, we investigate four distinct final states, $\gamma\gamma\gamma$, $\gamma\gamma Z$, $\gamma ZZ$, and $ZZZ$, which together provide complementary probes of neutral gauge boson interactions. Constraints are derived for the dimension-eight SMEFT operators parameterized by the Wilson coefficients $f_{T_0}$ through $f_{T_9}$.

The analysis is carried out at center-of-mass energies of $\sqrt{s} = 3\text{ TeV}$ with an integrated luminosity of $1\text{ ab}^{-1}$ and $\sqrt{s} = 10\text{ TeV}$ with $10\text{ ab}^{-1}$. As demonstrated in the cut-flow for cross~sections summaries for the various channels (Tables~\ref{tab:threephotoncutflow}, \ref{tab:twophotoncutflow}, \ref{tab:onephotoncutflow}, and~\ref{tab:nophotoncutflow}), the optimized event selections established in the preceding subsections effectively suppress the non-resonant SM backgrounds while maintaining high signal efficiency.

To translate these final event yields into physical limits, we employ a standard $\chi^2$ statistical framework. The signal cross~section's dependence on any given Wilson coefficient $f_{T_i}$ is parameterized quadratically as:
\begin{equation}
S(f_{T_i}) = S_{\text{SM}} + f_{T_i} S_{\text{int}} + {f_{T_i}}^2 S_{\text{EFT}},
\end{equation}
where $S_{\text{SM}}$, $S_{\text{int}}$, and $S_{\text{EFT}}$ denote the SM expectation, the interference term, and the pure EFT quadratic contribution, respectively. Using the expected event counts in our final signal regions, we define the test statistic as~\cite{Barlow:1993dm,Cowan:2010js}:
\begin{equation}
\chi^2(f_{T_i}) = \frac{\left[S(f_{T_i}) - S_{\text{SM}}\right]^2}{B + (\delta_{\text{sys}} B)^2},
\end{equation}
where $B$ represents the total background yield, and $\delta_{\text{sys}}$ accounts for fractional systematic uncertainties. To account for potential experimental imperfections, luminosity determination, and theoretical cross~section uncertainties, a conservative overall systematic uncertainty of $5\%$ is included. Assuming the SM background-only hypothesis, expected one-dimensional 95\% confidence level (CL) limits on the Wilson coefficients are established by the condition $\Delta\chi^2(c_i) = \chi^2(c_i) - \chi^2_{\text{min}} \le 3.841$, corresponding to one degree of freedom.

The four final states are evaluated independently across both center-of-mass energy stages, with the resulting 95\% CL bounds on the Wilson coefficients compiled in Tables~\ref{result1}--\ref{result4}. As shown in these constraint tables, the projected limits strengthen significantly with increasing collision energy across all evaluated processes. Furthermore, as expected, channels involving leptonic $Z$-boson decays yield somewhat weaker bounds due to the associated branching ratio suppression. Notably, the $f_{T_8}$ and $f_{T_9}$ coefficients consistently produce the most stringent constraints among all considered dimension-eight interactions, largely owing to their enhanced signal cross~sections. Another highlight of this study is that, to the best of our knowledge, the sensitivities to the $f_{T_3}$ and $f_{T_4}$ coefficients are established for the first time in the context of neutral triboson production.

\begin{table}[h!]
\centering
\caption{Expected 95\% CL bounds on the Wilson coefficients for the $\mu^+\mu^- \to \gamma\gamma\gamma$ process at $\sqrt{s} = 3$ and $10~\mathrm{TeV}$, assuming a $5\%$ systematic uncertainty.}
\label{result1}
\begin{tabular}{|c|c|c|}
\hline
& \multicolumn{1}{c|}{$\sqrt{s}=3 \ \ \mathrm{TeV}, (10^{-2} \ \ \mathrm{TeV^{-4}}), 1 \ \ \mathrm{ab^{-1}}$}
& \multicolumn{1}{c|}{$\sqrt{s}=10 \ \ \mathrm{TeV}, (10^{-4} \ \ \mathrm{TeV^{-4}}), 10 \ \ \mathrm{ab^{-1}}$} \\
\hline
$f_{T_0}/\Lambda^4$
& $[-65.29, 41.79]$
& $[-23.71, 39.98]$\\
$f_{T_1}/\Lambda^4$
& $[-65.35, 41.85]$
& $[-23.73, 40.00]$ \\
$f_{T_2}/\Lambda^4$
& $[-168.61, 86.31]$
& $[-85.44, 43.95]$ \\
$f_{T_3}/\Lambda^4$
& $[-168.5, 86.20]$
& $[-85.38, 43.89]$ \\
$f_{T_4}/\Lambda^4$
& $[-33.48, 8.73]$
& $[-3.732, 4.584]$  \\
$f_{T_5}/\Lambda^4$
& $[-12.89, 4.55]$
& $[-0.693, 0.709]$ \\
$f_{T_6}/\Lambda^4$
& $[-12.88, 4.55]$ 
& $[-0.692, 0.708]$ \\
$f_{T_7}/\Lambda^4$
& $[-33.47, 8.73]$
& $[-3.728, 4.575]$ \\
$f_{T_8}/\Lambda^4$
& $[-0.540, 0.395]$
& $[-0.0264,0.0264]$ \\
$f_{T_9}/\Lambda^4$
& $[-4.976, 1.074]$
& $[-0.0573,0.0573]$\\

\hline

\end{tabular}
\end{table}

\begin{table}[h!]
\centering
\caption{Expected 95\% CL bounds on the Wilson coefficients for the $\mu^+\mu^- \to \gamma\gamma Z$ process at $\sqrt{s} = 3$ and $10~\mathrm{TeV}$, assuming a $5\%$ systematic uncertainty.}
\label{result2}
\begin{tabular}{|c|c|c|}
\hline
& \multicolumn{1}{c|}{$\sqrt{s}=3 \ \ \mathrm{TeV}, (10^{-2} \ \ \mathrm{TeV^{-4}}), 1 \ \ \mathrm{ab^{-1}}$}
& \multicolumn{1}{c|}{$\sqrt{s}=10 \ \ \mathrm{TeV}, (10^{-4} \ \ \mathrm{TeV^{-4}}), 10 \ \ \mathrm{ab^{-1}}$} \\
\hline
$f_{T_0}/\Lambda^4$
& $[-31.41, 20.46]$
& $[-70.24, 50.94]$ \\
$f_{T_1}/\Lambda^4$
& $[-31.42, 20.47]$
& $[-70.18, 50.88]$ \\
$f_{T_2}/\Lambda^4$
& $[-71.15, 44.59]$
& $[-125.47, 119.76]$ \\
$f_{T_3}/\Lambda^4$
& $[-71.21, 44.64]$
& $[-125.48, 119.77]$ \\
$f_{T_4}/\Lambda^4$
& $[-29.52, 29.86]$
& $[-101.04, 47.40]$ \\
$f_{T_5}/\Lambda^4$
& $[-11.16, 10.580]$
& $[-9.804, 14.414]$ \\
$f_{T_6}/\Lambda^4$
& $[-14.31, 14.13]$
& $[-19.60, 63.01]$ \\
$f_{T_7}/\Lambda^4$
& $[-35.07, 33.97]$
& $[-58.23, 103.50]$ \\
$f_{T_8}/\Lambda^4$
& $[-3.024, 2.166]$
& $[-1.338, 1.460]$ \\
$f_{T_9}/\Lambda^4$
& $[-6.262, 4.311]$
& $[-17.34, 10.92]$ \\
\hline
\end{tabular}
\end{table}

\begin{table}[h!]
\centering
\caption{Expected 95\% CL bounds on the Wilson coefficients for the $\mu^+\mu^- \to \gamma Z Z$ process at $\sqrt{s} = 3$ and $10~\mathrm{TeV}$, assuming a $5\%$ systematic uncertainty.}
\label{result3}
\begin{tabular}{|c|c|c|}
\hline
& \multicolumn{1}{c|}{$\sqrt{s}=3 \ \ \mathrm{TeV}, (10^{-2} \ \ \mathrm{TeV^{-4}}), 1 \ \ \mathrm{ab^{-1}}$}
& \multicolumn{1}{c|}{$\sqrt{s}=10 \ \ \mathrm{TeV}, (10^{-4} \ \ \mathrm{TeV^{-4}}), 10 \ \ \mathrm{ab^{-1}}$} \\
\hline
$f_{T_0}/\Lambda^4$
& $[-45.92, 41.36]$
& $[-267.16, 325.82]$ \\
$f_{T_1}/\Lambda^4$
& $[-45.90, 41.33]$
& $[-266.04, 324.70]$ \\
$f_{T_2}/\Lambda^4$
& $[-98.11, 82.99]$
& $[-595.88, 636.52]$ \\
$f_{T_3}/\Lambda^4$
& $[-98.36, 83.25]$
& $[-593.46, 634.10]$ \\
$f_{T_4}/\Lambda^4$
& $[-96.95, 100.52]$
& $[-644.48, 692.79]$ \\
$f_{T_5}/\Lambda^4$
& $[-30.84, 33.34]$
& $[-202.64, 241.24]$ \\
$f_{T_6}/\Lambda^4$
& $[-55.31, 56.09]$
& $[-355.44, 424.07]$ \\
$f_{T_7}/\Lambda^4$
& $[-136.92, 146.22]$
& $[-915.26, 977.24]$ \\
$f_{T_8}/\Lambda^4$
& $[-16.31, 15.35]$
& $[-98.20,120.15]$ \\
$f_{T_9}/\Lambda^4$
& $[-35.00, 30.06]$
& $[-196.60, 257.05]$ \\
\hline
\end{tabular}
\end{table}

\begin{table}[h!]
\centering
\caption{Expected 95\% CL bounds on the Wilson coefficients for the $\mu^+\mu^- \to  Z Z Z$ process at $\sqrt{s} = 3$, assuming a $5\%$ systematic uncertainty.}
\label{result4}
\begin{tabular}{|c|c|}
\hline
& \multicolumn{1}{c|}{$\sqrt{s}=3 \ \ \mathrm{TeV}, (10^{-2} \ \ \mathrm{TeV^{-4}}), 1 \ \ \mathrm{ab^{-1}}$} \\
\hline
$f_{T_0}/\Lambda^4$
& $[-137.98, 135.56]$ \\
$f_{T_1}/\Lambda^4$
& $[-137.38, 134.96]$ \\
$f_{T_2}/\Lambda^4$
& $[-283.94, 278.72]$ \\
$f_{T_3}/\Lambda^4$
& $[-286.14, 281.21]$ \\
$f_{T_4}/\Lambda^4$
& $[-470.06, 463.42]$ \\
$f_{T_5}/\Lambda^4$
& $[-225.19, 222.80]$ \\
$f_{T_6}/\Lambda^4$
& $[-225.49, 223.09]$ \\
$f_{T_7}/\Lambda^4$
& $[-469.15, 462.51]$ \\
$f_{T_8}/\Lambda^4$
& $[-163.40, 163.36]$ \\
$f_{T_9}/\Lambda^4$
& $[-337.60, 337.65]$ \\
\hline
\end{tabular}
\end{table}

\section{Conclusion} \label{sec:summary}                                       

Anomalous quartic gauge couplings (aQGCs) offer a sensitive probe of beyond the Standard Model (BSM) physics, as subtle deviations in these interactions can encode the effects of heavy new states beyond the direct kinematic reach of current colliders. Neutral quartic gauge interactions are particularly compelling testing grounds, as the corresponding vertices are strictly absent at tree level in the Standard Model (SM). Triboson production provides a direct experimental probe of these couplings, offering complementary sensitivity to anomalous gauge interactions. In this context, a high-energy muon collider represents an ideal facility, combining multi-TeV collision energies with the clean environment of leptonic initial states and negligible QCD backgrounds relative to hadron colliders.

In this work, we have performed a comprehensive phenomenological study of neutral aQGCs in the triboson final states $\mu^+\mu^- \to \gamma\gamma\gamma$, $\gamma\gamma Z$, $\gamma ZZ$, and $ZZZ$ within the Standard Model Effective Field Theory (SMEFT) framework. We focused on the ten dimension-eight transverse operators ($\mathcal{O}_{T_0}$--$\mathcal{O}_{T_9}$). The analysis was conducted at prospective center-of-mass energies of $\sqrt{s} = 3\text{ TeV}$ and $10\text{ TeV}$, assuming integrated luminosities of $1\text{ ab}^{-1}$ and $10\text{ ab}^{-1}$, respectively. Following optimized cut-based event selections designed to maximize signal purity, we extracted projected 95\% confidence level (CL) bounds on the Wilson coefficients using a $\chi^2$ test statistic incorporating realistic systematic uncertainties. A detailed treatment of beam-induced backgrounds (BIB) and full detector reconstruction--while critical for future experimental implementation--is beyond the scope of this phenomenological study. 

Our findings demonstrate that the sensitivity to neutral aQGCs varies substantially across the different final states and operator Lorentz structures. Among the evaluated channels, the all-photon final state ($\gamma\gamma\gamma$) delivers the most stringent bounds---particularly on the $f_{T_8}/\Lambda^4$ and $f_{T_9}/\Lambda^4$ coefficients---owing to its high reconstruction efficiency and clean signal topology. Conversely, channels containing $Z$ bosons yield somewhat looser bounds due to branching ratio suppressions from leptonic $Z$ decays. Crucially, this study establishes novel constraints on the $f_{T_3}/\Lambda^4$ and $f_{T_4}/\Lambda^4$ coefficients, which, to the best of our knowledge, have not been previously evaluated in these neutral triboson modes. 

Finally, direct comparisons of our $f_{T_8}$ and $f_{T_9}$ limits with previous literature are detailed in Tables~\ref{tab:T8_comparison} and~\ref{tab:T9_comparison}, demonstrating the substantial gain in reach enabled by multi-TeV muon collisions. For reference, the tables also include the direct LHC constraints on these coefficients from neutral triboson production, obtained from $\gamma\gamma Z$ measurements at $\sqrt s= 13\text{ TeV}$ by the CMS \cite{CMS:2021jji} and ATLAS \cite{ATLAS:2022wmu} collaborations, where the tables report the CMS limits, which are the tighter of the two.  As illustrated in these tables, our constraint derived from the triphoton channel ($\gamma\gamma\gamma$) is competitive with previous projections at a muon collider. Furthermore, our bound in the $\gamma\gamma Z$ final state is comparable to those expected from the FCC-hh while being significantly stronger than the projected limits from the HL-LHC and HE-LHC.

\begin{table*}[t]
    \centering
    \caption{Comparison of the constraints on the anomalous coupling parameter $f_{T_8}/\Lambda^4$ obtained 
    in this work across various neutral triboson channels with existing limits from previous studies and collider facilities.}
    \label{tab:T8_comparison}
    \scriptsize
    \setlength{\tabcolsep}{4pt}
    \renewcommand{\arraystretch}{1.05}
    \resizebox{\textwidth}{!}{%
    \begin{tabular}{l c c c c c}
        \toprule
        \textbf{Collider} &
        \textbf{$\sqrt{s}$} &
        \textbf{Luminosity} &
        \textbf{Sys. Error} &
        \textbf{95\% C.L. interval on $f_{T_8}/\Lambda^4 (\mathrm{TeV^{-4}})$} &
        \textbf{Ref.} \\
        \midrule
        CLIC & $3~\mathrm{TeV}$ & $5~\mathrm{ab}^{-1}$ & $5\%$
        & $[-1.96,2.03]\times 10^{-2}$ & \citep{Ari:2021rmx} \\

        CLIC (Considering ISR and Beamstrahlung)
        & $3~\mathrm{TeV}$ & $5~\mathrm{ab}^{-1}$ & $ 5 \%$
        & $[-2.13,2.50]\times 10^{-2}$ & \citep{Ari:2021rmx} \\

        Muon Collider (Cut-and-count)
        & $30~\mathrm{TeV}$ & $10~\mathrm{ab}^{-1}$ & $S=2$
        & $[-0.233,0.102]\times 10^{-5}$ & \citep{Yang:2020rjt} \\

        Muon Collider (KMAD)
        & $30~\mathrm{TeV}$ & $10~\mathrm{ab}^{-1}$ & $S=2$
        & $[-0.195,0.0890]\times 10^{-5}$ & \citep{Zhang:2023yfg} \\

        Muon Collider (PCAAD)
        & $30~\mathrm{TeV}$ & $10~\mathrm{ab}^{-1}$ & $S=2$
        & $[-0.213,0.0991]\times 10^{-5}$ & \citep{Dong:2023nir} \\
        
        HL-LHC & $14~\mathrm{TeV}$ & $3~\mathrm{ab}^{-1}$ & $ 5 \%$
        & $ [-2.35,2.26] \times 10^{-1}$ & \citep{Senol:2021wza} \\

        HE-LHC & $27~\mathrm{TeV}$ & $15~\mathrm{ab}^{-1}$ & $ 5 \%$
        & $ [-4.44,3.47] \times 10^{-2}$ & \citep{Senol:2021wza} \\

        FCC-hh & $100~\mathrm{TeV}$ & $30~\mathrm{ab}^{-1}$ & $ 5 \%$
        & $ [-1.50,0.89] \times 10^{-3}$ & \citep{Senol:2021wza} \\
        
        FCC-hh & $100~\mathrm{TeV}$ & $30~\mathrm{ab}^{-1}$ & $ 0 \%$
        & $ [-4.01,4.01] \times 10^{-5}$ & \citep{Senol:2023fpa} \\
        
        FCC-hh (BDT) & $100~\mathrm{TeV}$ & $30~\mathrm{ab}^{-1}$ & $ 5 \%$
        & $ [-4.07, 4.07] \times 10^{-3}$ & \citep{Yilmaz:2026afl} \\
        
        LHC (CMS) & $13~\mathrm{TeV}$ & $137~\mathrm{fb}^{-1}$ & $ - $
        & $ [-1.06, 1.10] $ & \citep{CMS:2021jji} \\

        \midrule
        \textbf{Our result ($ \gamma \gamma \gamma$)}
        & $10~\mathrm{TeV}$ & $10~\mathrm{ab}^{-1}$ & $5\%$
        & $\boldsymbol{[-2.64,2.64] \times 10^{-6}}$ &  \\

        \textbf{Our result ($ \gamma \gamma Z $)}
        & $10~\mathrm{TeV}$ & $10~\mathrm{ab}^{-1}$ & $5\%$
        & $\boldsymbol{[-1.34, 1.46] \times 10^{-4}}$ &  \\
        
        \textbf{Our result ($ \gamma Z Z $)}
        & $10~\mathrm{TeV}$ & $10~\mathrm{ab}^{-1}$ & $5\%$
        & $\boldsymbol{[-9.82,12.01] \times 10^{-3}}$ &  \\
        
        \textbf{Our result ($ Z Z Z $)}
        & $ 3 ~\mathrm{TeV}$ & $1~\mathrm{ab}^{-1}$ & $5\%$
        & $\boldsymbol{[-163.40, 163.36]}$ &  \\
        
        \bottomrule
    \end{tabular}%
    }
\end{table*}

\begin{table*}[t]
    \centering
    \caption{Comparison of the constraints on the anomalous coupling parameter $f_{T_9}/\Lambda^4$ obtained 
    in this work across various neutral triboson channels with existing limits from previous studies and collider facilities.}
    \label{tab:T9_comparison}
    \scriptsize
    \setlength{\tabcolsep}{4pt}
    \renewcommand{\arraystretch}{1.05}
    \resizebox{\textwidth}{!}{%
    \begin{tabular}{l c c c c c}
        \toprule
        \textbf{Collider} &
        \textbf{$\sqrt{s}$} &
        \textbf{Luminosity} &
        \textbf{Sys. Error} &
        \textbf{95\% C.L. interval on $f_{T_9}/\Lambda^4 (\mathrm{TeV^{-4}})$} &
        \textbf{Ref.} \\
        \midrule
        CLIC & $3~\mathrm{TeV}$ & $5~\mathrm{ab}^{-1}$ & $5\%$
        & $[-4.20,3.85]\times 10^{-2}$ & \citep{Ari:2021rmx} \\

        CLIC (Considering ISR and Beamstrahlung)
        & $3~\mathrm{TeV}$ & $5~\mathrm{ab}^{-1}$ & $ 5 \%$
        & $[-4.65,4.61]\times 10^{-2}$ & \citep{Ari:2021rmx} \\

        Muon Collider (Cut-and-count)
        & $30~\mathrm{TeV}$ & $10~\mathrm{ab}^{-1}$ & $S=2$
        & $[-0.553, 0.167]\times 10^{-5}$ & \citep{Yang:2020rjt} \\

        Muon Collider (KMAD)
        & $30~\mathrm{TeV}$ & $10~\mathrm{ab}^{-1}$ & $S=2$
        & $[-0.458, 0.146]\times 10^{-5}$ & \citep{Zhang:2023yfg} \\

        Muon Collider (PCAAD)
        & $30~\mathrm{TeV}$ & $10~\mathrm{ab}^{-1}$ & $S=2$
        & $[-0.503,0.168]\times 10^{-5}$ & \citep{Dong:2023nir} \\

        HL-LHC & $14~\mathrm{TeV}$ & $3~\mathrm{ab}^{-1}$ & $ 5 \%$
        & $ [-4.76,4.40] \times 10^{-1}$ & \citep{Senol:2021wza} \\

        HE-LHC & $27~\mathrm{TeV}$ & $15~\mathrm{ab}^{-1}$ & $ 5 \%$
        & $ [-8.11,6.11] \times 10^{-2}$ & \citep{Senol:2021wza} \\

        FCC-hh & $100~\mathrm{TeV}$ & $30~\mathrm{ab}^{-1}$ & $ 5 \%$
        & $ [-1.48,1.31] \times 10^{-3}$ & \citep{Senol:2021wza} \\
        
        FCC-hh & $100~\mathrm{TeV}$ & $30~\mathrm{ab}^{-1}$ & $ 0 \%$
        & $ [-2.20,2.20] \times 10^{-4}$ & \citep{Senol:2023fpa} \\
        
        FCC-hh (BDT) & $100~\mathrm{TeV}$ & $30~\mathrm{ab}^{-1}$ & $ 5 \%$
        & $ [-9.16, 9.16] \times 10^{-3}$ & \citep{Yilmaz:2026afl} \\
        
        LHC (CMS) & $13~\mathrm{TeV}$ & $137~\mathrm{fb}^{-1}$ & $ - $
        & $ [-1.82, 1.82] $ & \citep{CMS:2021jji} \\

        \midrule
        \textbf{Our result ($ \gamma \gamma \gamma$)}
        & $10~\mathrm{TeV}$ & $10~\mathrm{ab}^{-1}$ & $5\%$
        & $\boldsymbol{[-5.73,5.73]\times 10^{-6}}$ &  \\

        \textbf{Our result ($ \gamma \gamma Z $)}
        & $10~\mathrm{TeV}$ & $10~\mathrm{ab}^{-1}$ & $5\%$
        & $\boldsymbol{[-1.73, 1.09] \times 10^{-3}}$ &  \\
        
        \textbf{Our result ($ \gamma Z Z $)}
        & $10~\mathrm{TeV}$ & $10~\mathrm{ab}^{-1}$ & $5\%$
        & $\boldsymbol{[-1.97, 2.57]\times 10^{-2}}$ &  \\
        
        \textbf{Our result ($ Z Z Z $)}
        & $3~\mathrm{TeV}$ & $1~\mathrm{ab}^{-1}$ & $5\%$
        & $\boldsymbol{[-337.60, 337.65]}$ &  \\
        
        \bottomrule
    \end{tabular}%
    }
\end{table*}

	\section*{Acknowledgments} \label{sec:ack}                                            
   The authors acknowledge the use of Google Gemini for AI-assisted language editing and stylistic refinement of the text. 
   All technical calculations, numerical analyses, and scientific conclusions remain solely the responsibility of the authors.

	\bibliographystyle{JHEP}                                                                                      
	\bibliography{Dark}

\end{document}